\documentclass[10pt,a4paper]{article}

\usepackage{graphicx,amsmath,amssymb,fixmath,color,isomath,theorem}
\usepackage{graphicx}

\usepackage{pgf,pgfarrows,pgfnodes,tikz,color,tkz-berge}
\usetikzlibrary{arrows,shapes,snakes,automata,backgrounds,petri,topaths,calc}

\newcommand{\floor}[1]{\lfloor #1 \rfloor}

\newcommand{\Ord}[1]{ {\mathcal O}( #1 )}

\newcommand{\vect}[1]{\ensuremath{ \mathbold #1 } }

\newtheorem{theorem}{Theorem}[section]
\newtheorem{lemma}[theorem]{Lemma}

\theoremstyle{definition}

\theoremstyle{remark}

\numberwithin{equation}{section}

\begin{document}

\title{Maximal switchability of centralized  networks}


\author{Sergei Vakulenko$^{1,2}$, Ivan Morozov$^3$ and Ovidiu Radulescu$^3$,  \\
\footnotesize  $^1$ Institute for Mech. Engineering Problems, Saint Petersburg, Russia \\
\footnotesize  $^2$  Saint Petersburg National Research University of Information Technologies, Mechanics and Optics, \\ \footnotesize Saint Petersburg, Russia \\
\footnotesize  $^3$ University of Technology and Design, St.Petersburg, Russia, \\
\footnotesize  $^4$ DIMNP UMR CNRS 5235, University of Montpellier 2, Montpellier, France.
 }

\maketitle

\centerline{\bf Abstract}

We consider continuous time Hopfield-like recurrent networks as
dynamical models for gene regulation and neural networks.
We are interested in networks that contain $n$ high-degree nodes preferably
connected to a large number of $N_s$ weakly connected satellites,
a property that we call $n/N_s$-centrality.
If  the hub dynamics is slow, we obtain that the large time network dynamics is completely
defined by the hub dynamics.
Moreover, such networks are maximally flexible and switchable,
in the sense that they can switch from a globally attractive rest state to any structurally stable
dynamics when the response time of a special controller hub is changed.
In particular, we show that a decrease of the controller hub response time
can lead to a sharp variation in the network attractor structure: we can obtain
a set of new local attractors, whose number can increase exponentially
with {$N$, the total number of nodes of the nework}.
These new attractors can be periodic or even chaotic.
We provide an algorithm, which allows us to design networks with the desired
switching properties, or to learn them from time series,
by adjusting the interactions between hubs and satellites.
Such switchable networks could be used as models for context dependent
adaptation in functional genetics or as models for cognitive functions
in neuroscience.


\tikzstyle{every picture}+=[remember picture]
\tikzstyle{na} = [baseline=-.5ex]

\section{Introduction}

Networks of dynamically coupled elements have imposed themselves as models of complex systems
in physics, chemistry, biology and engineering \cite{newman2003structure}.
The most studied propriety of networks is { their topological structure}. Structural features
of networks are usually defined by the distribution of
the number of direct connections a node has, or by various statistical properties of
paths and circuits in the network \cite{newman2003structure,AB}.
An important structure related property of networks is their scale-freeness \cite{Jeong1,Jeong2, AB, Bas1}
often invoked as a paradigm of self-organization and spontaneous emergence of complex
collective behaviour \cite{chialvo}.
In scale-free networks  the fraction $P(k)$ of nodes in the network having $k$ connections
to other nodes (i.e. having degree $k$) can be estimated for large values of $k$ as
$P(k) \ \sim \ k^{-\gamma}$, where $ \gamma$  is a parameter whose value is typically in the range $2 < \gamma < 3$ \cite{AB}.
In such networks, the degree is extremely heterogeneous. In particular, there are
strongly connected nodes that can be named hubs, or centers.
The hubs communicate to each other directly, or via a number of weakly connected nodes.
The weakly connected nodes that interact mainly with hubs can be called satellites.
Scale-free networks have also nodes of intermediate connectivity.
Networks that have only two types of nodes, strongly connected hubs and weakly connected
satellites  are known as bimodal degree networks \cite{tanizawa2005optimization}.
Because of the presence of a large number of hubs, scale-free or bimodal degree networks
can be called centralized.
Centralized  connectivity has been {found} by functional imaging of brain activity
in neuroscience \cite{chialvo}, and also by large scale studies of the
protein-protein interactions or of the metabolic networks in functional genetics \cite{Jeong1,Jeong2}.

The centralized architecture was shown to be important for many emergent properties
of networks. For instance, there has been a lot of interest in the resilience of networks with
respect to attacks that remove some of their components \cite{albert2000error}.
 It {was} shown that networks with bimodal degree connectivity are resilient to simultaneous
targeted and random attacks \cite{tanizawa2005optimization}, whereas scale-free networks are
robust with respect
to random attacks, but sensitive to targeted attacks that are directed against hubs
\cite{cohen2001breakdown,bar2004response}. For this reason, the term  "robust-yet-fragile"
was coined in relation to scale-free networks \cite{carlson2002complexity}.

From a more dynamical perspective, a centralized architecture facilitates
communication between hubs, stabilizes hubs by making them insensitive to noise
\cite{vakulenko2012flexible,vakulenko2011flexible}
and allows for hub synchronization even in the absence of satellite synchronization \cite{pereira2013connectivity,pereira2010hub,stroud2015dynamics}.
Another important question concerning networks is how to
push their dynamics from one region of the phase space to another or from one type of
behaviour to another, briefly how to {control the network dynamics} \cite{lin1974structural,sun2013controllability,nepusz2012controlling,cowan2012nodal,ruths2014control,
pasqualetti2014controllability,jia2013control,pan2014structural,gao2014target,yuan2013exact}. Several authors used
Kalman's results for linear systems to understand how network structure influences network {dynamics} controllability,
and in particular how to choose the control nodes \cite{lin1974structural,nepusz2012controlling,cowan2012nodal}.
As pointed out by \cite{motter2015networkcontrology,lai2014controlling} several difficulties occur when one tries to apply these general results to real networks.
Even for linear networks, the control of trajectories is nonlocal
\cite{sun2013controllability} and shortcuts are rarely allowed.
As a result, even small changes of the network state may ask for control
signals of large amplitude and energy \cite{yan2012controlling}.
The control of nonlinear networks is even more difficult and in this case  we have no general  results.
Nonlinear networks can have several co-existing attractors and it is interesting to find
out how to push the state of the network from one attractor basin to another.
The ability of networks to change attractor under the effect
of targeted perturbations can be called switchability.
In relation to this, the paper  \cite{SyS} has introduced the
terminology "stable yet switchable" (SyS) meaning that the network
remains stable given a context and is able  to reach another stable state
when a stimulus indicates a change of the context.  It was shown, by numerical simulations, that
centralized networks with bimodal degree distribution are more prone to SyS behavior than
scale-free networks \cite{SyS}.
Switchability is important for practical reasons, for instance in drug design.
In such applications, one uses pharmaceutical action on nodes to push  a network that functions in a
pathological attractor (such pathological attractors were discussed in relation to cancer
\cite{Huang2009869} or neurological disorders \cite{stam1995investigation,edwards1999parkinsonian})
to a healthy functioning mode, characterized by a different  attractor.
Numerical methods to study switchability of linear \cite{wu2014transittability} and nonlinear \cite{cornelius2013realistic} networks were discussed  in relation with drug design in cancer research.
In theoretical biology, network
switchability can be important for mathematical theories of genetic adaptation \cite{orr2005genetic}.
If one looks at organisms as complex systems and model them by networks, then
 adaptation to changes in the environment can be described as switching the network
from one attractor to another one with a higher fitness \cite{orr2005genetic}.
An important question that is often asked with respect to{ tuning network dynamics} is how many
driver nodes are needed to control {that dynamics}. For linear networks, it was shown that this number
is large if {we  aim to obtain a total control}, which allows us to switch the network between any pair of states. This number
can be as high as $80\%$ for molecular regulatory networks \cite{liu2011controllability}. This fact, as emphasized in
\cite{wu2014transittability}, contradicts empirical results about cellular reprogramming and about
adaptive evolution. Much less nodes are
needed if instead of full controlability one wants switching between specific pairs of unexpected and
desired states \cite{wu2014transittability}. This concept, named ``transittability'' in  \cite{wu2014transittability}, {is very similar to} our switchability, but was studied only for linear systems.

In this paper, we study dynamical properties of large nonlinear networks with centralized architecture.
We consider continuous time versions of the Hopfield model of recurrent neural networks \cite{Hopfield}
with a large number $N$ of neurons. The Hopfield model is based on the two-states McCullogh and
Pitts formal neuron and uses symmetrical weight matrices to specify interactions between neurons.
Like to the Hopfield version, we use
a thresholding function to describe switching between the two neuron states, active and inactive.
However, contrary to the original Hopfield version, we do not impose symmetrical interactions between
neurons, {in other words} our weight matrix is not necessarily symmetric.
 This model has been successfully used to describe associative memories \cite{Hopfield}, neural computation \cite{hopfield1986computing,maass1991computational}, disordered systems in statistical physics \cite{talagrand1998rigorous}, neural activity  \cite{li1989modeling,edwards1999parkinsonian} and also to investigate space-time dynamics of gene networks in molecular biology \cite{Rein1,vohradsky2001neural}.
The choice of such type of dynamics is motivated by the existence of universal approximation results for multilayered perceptrons (see, for example, \cite{Barron}). In particular, we have shown elsewhere that networks with Hopfield-type dynamics can approximate
any structurally stable dynamics, including reaction-diffusion biochemical networks also largely
used in biology \cite{vakulenko2011flexible}.

Our aim is to study analytically the ability of a network with centralized
architecture to be switchable. We employ a special notion of centrality.
Many biological networks exhibit so-called dissortative mixing, i.e.,  high-degree nodes
are preferably connected to low-degree nodes \cite{Lip}.
We   will consider networks with $n$ strongly connected hubs.
We also assume that each hub is under the action of at least $N_s$
weakly connected satellites, that on turn receive actions from all the hubs.
For large networks, $N_s$ increases at least as fast as a power of $N$, $N_s > c_0  N^{\theta}$
where $c_0 > 0$, $0 < \theta < 1$ are constants and $N$ is the total number of nodes.
We call this property $n/N_s$-centrality.
This network architecture ensures a large number of feed-back loops
that produce complex dynamics. Furthermore, the dissortative connectivity implies functional
heterogeneity of the hubs and satellites. The hubs play the role of controllers
and the satellites sustain the feedback loops needed for attractor multiplicity. The large  number of satellites {guarantees a sufficient flexibility of the
network dynamics} and also buffer the perturbations transmitted to the hubs. This principle
applies well to gene networks. The hubs in such networks
can be the transcription factors, which are stabilized by numerous interactions
 with non-coding RNAs that represent the satellites  \cite{Li}.
In addition to structural conditions, we will consider a special correlation between
time scales and connectivity of the nodes: the hubs have slow response, whereas the satellites
respond rapidly. This condition is natural for {many} real networks. { The hubs have to cope with
multiple tasks, therefore they must have more complex interaction than the satellites. Consequently, the hubs
need more resources to be {produced, decomposed, and} react with other nodes, therefore their dynamics is  slow}.
This property is obvious for gene networks, where transcription {factors} are complex
proteins, much {larger and} more stable than the non-coding RNAs.

Our first result is valid without conditions on the structure and depends
only on the condition on the timescales. We assume that there exist $n << N$ slow nodes,  whereas
all the remaining ones are fast. {Then,} the dynamics of the network can be reduced to $n$ variables.
We prove the existence of an inertial manifold of dimension $n$, which completely  captures all
network dynamics for large times. We recall that the fundamental  concept of inertial manifold
was introduced for  infinite dimensional and multidimensional systems.
The inertial manifolds are globally attracting invariant ones \cite{Temam}. The large time dynamics of a
system possessing an inertial  manifold, is defined by a smooth vector field $F$ of relatively small
dimension,  so-called inertial form. All attractors lie on  inertial  manifold \cite{Temam}.

The second result holds under the structural assumption that the network
is $n/N_s$-central.  Under this condition, we show that the inertial forms $F$
obtained from such networks are dense in the set of all smooth vector fields of
dimension $n$. This implies that given a certain combination of attractors defined by
vector fields $Q_i$ we can construct a centralized network that exhibits a combination of attractors that is
topologically equivalent to the one given.
Furthermore, we show that $n/N_s$-central networks  can exhibit
''maximal switchability". By changing a control parameter $\xi$, which determines
the response time of a single  network hub (''controller" hub), we can sharply change the
network attractor. For instance we can switch from a
situation when the network has a single rest point for $\xi > \xi_0$
to a situation when the network has a complicated global attractor for $\xi < \xi_0$,
including a number of  local attractors, which may be periodic or chaotic.
The network state tends to the corresponding local attractor depending on
the initial state of the control hub.
This result shows in an analytical and rigorous way how nonlinear networks can be
switched by only one control node. The possibility of switching nonlinear networks
by a small number of nodes is crucial in theories of
genetic adaptation. Indeed, phenomenological theories predict and empirical data confirm
that {the main part} of the adaptive evolution process
consists in only a few mutations producing large fitness changes \cite{orr2005genetic}.

Our third result proves, in an analytical way, that the number
of rest point local attractors (and therefore the network capacity) of $n/N_s$-central
networks may be exponentially large in the number of nodes.

We also describe a constructive algorithm, which allows us
to obtain a centralized network
that performs a prescribed inertial dynamics and the desired
switching properties of the network.

\section{Problem statement  and main assumptions}

We consider the Hopfield-{like} networks \cite{Hopfield} described by the ordinary differential equations
\begin{equation} \label{Hop1}
\frac{du_i}{dt}=\sigma( \sum_{j=1}^N W_{ij} u_j -h_i)  -  \lambda_i  u_i,
\end{equation}
where $u_i$, $h_i$ and $\lambda_i >0$, $i=1,..., N$ are node activities,
activation thresholds and degradation coefficients, respectively.
The matrix entry $W_{ij}$ describes the action
of the node $j$ on the node $i$, which is an activation  if $W_{ij}>0$
or a repression if $W_{ij}<0$. Contrary to the original Hopfield model, the interaction
matrix $W$ is not necessarily symmetric.
The function $\sigma$ is an increasing  and smooth (at least twice differentiable) "sigmoidal" function such
that
\begin{equation}
\sigma(-\infty)=0, \quad \sigma(+\infty)=1, \quad \sigma^{'}(z) >0.
\label {eq2.5}
\end{equation}
 Typical examples can be given
by
\begin{equation}
     \sigma(h)=\frac{1}{1 + \exp(- h)}, \quad \sigma(h)
     = \frac{1}{2} \left( \frac{h}{\sqrt{1+h^2}} + 1 \right).
\label {eq2.6}
\end{equation}

The structure of interactions in the model is defined by a weighted digraph $(V, E, W)$ with the  set $V$ of nodes, the edge set $E$ and weights $W_{ij}$.
The nodes $v_j$, $j=1 ..., N$ can be neurons or genes, depending on applications.

{\bf Assumption 1.}

{\em Assume that  if
$W_{ji}\ne 0$,  then $(i,j)$ is an edge of the graph,  $(i, j) \in E$.  This means that the
 $i$-th {node} can act on the $j$-th node only if it is prescribed by an edge of the digraph $(V, E, W)$.  We also suppose that $(i, i) \notin E$, i.e., the nodes
do not act on themselves}.

Assume that  the digraph $(V, E, W)$ satisfies a condition, which is a variant of the centrality property.  This condition is a purely topological  one and
thus it is independent on the weights $W_{ij}$.
To formulate this condition, we introduce a special notation.

Let us consider a node $v_j$.  Let us denote by $S^*(j)$ the set of all nodes, which act on the neuron $j$:
\begin{equation}
S^*(j)=\{ v_i \in V:     \quad     edge \ (i, j) \in  E  \}.
\end{equation}
For each set of nodes ${\mathcal C} \subset V$ we introduce the set ${\mathcal S}({\mathcal C})$ of the nodes, which are under action of all nodes from ${\mathcal C}$ and
which are not belonging to ${\mathcal C}$:
\begin{equation} \label{centrality}
{\mathcal S}({\mathcal C})=\{v_i \in V:  \ for \ each \ j\in {\mathcal C} \ edge \ (j, i) \in E \  and  \ v_i \notin  {\mathcal C} \}.
\end{equation}

\vspace{0.2cm}

{\bf $n/N_s$-Centrality assumption.}
{\em
 The graph $(V, E, W)$ is connected and there exists a set of  nodes  ${\mathcal C}$  such that

{\bf i} ${\mathcal C}$ consists of $n$ nodes;

{\bf ii} for each $j \in {\mathcal C}$ the  intersection  ${\mathcal S}^*(j)  \cap {\mathcal S}({\mathcal C})$ contains at least $N_s$ nodes, where $N_s > c_0 N^{\theta}$
with constants  $c_0>0,  \theta \in (0,1)$, which are independent of $j$ and $N$.
}
\vspace{0.2cm}

 The nodes from ${\mathcal C}$  can be interpreted {as} hubs (centers) and the nodes from   ${\mathcal S}({\mathcal C})$
  {are the} satellites.
 The condition {\bf ii} implies that each center is under action of sufficiently many satellites. In turn, if we consider the union of these
satellites, all the centers act on them (see Fig.\ref{fig:1}). Such an intensive interaction {leads}, as we will see below, to a very complicated large time behaviour.

\begin{figure}[h!]
\begin{center}
\scalebox{1}{
\begin{tikzpicture}
 \SetUpEdge[lw         = 0.5pt,
            color      = black,
            labelstyle = {fill=white, sloped}]
  \tikzset{node distance = 2.5cm,main node/.style={circle,fill=blue!12,draw,font=\sffamily\large\bfseries}
   }
 \tikzstyle{vertex}=[circle,fill=black!25,minimum size=7pt,inner sep=0pt]
  \node[vertex, circle,xshift=-3cm] (w1) at ({360/3 * 0 }:2) {$w_1$};
 \node[vertex, circle,xshift=-3cm] (w2) at ({360/3 * 1 }:2) {$w_2$};
 \node[vertex, circle,xshift=-3cm] (w3) at ({360/3 * 2 }:2) {$w_3$};
 \node[draw, circle,xshift=-3cm] (v1) at (0:0) {$v_1$};
 \node[vertex, circle,xshift=3cm] (w4) at ({360/3 * 0+ 60}:2) {$w_4$};
 \node[vertex, circle,xshift=3cm] (w5) at ({360/3 * 1+ 60}:2) {$w_5$};
 \node[vertex, circle,xshift=3cm] (w6) at ({360/3 * 2+ 60}:2) {$w_6$};
 \node[draw, circle,xshift=3cm] (v2) at (0:0) {$v_2$};
 \tikzstyle{EdgeStyle}=[post,bend left,line width = 1]
 \Edge(v1)(w1)
  \Edge(w1)(v1)
   \Edge(v1)(w2)
  \Edge(w2)(v1)
   \Edge(v1)(w3)
  \Edge(w3)(v1)
   \Edge(v2)(w4)
  \Edge(w4)(v2)
   \Edge(v2)(w5)
  \Edge(w5)(v2)
   \Edge(v2)(w6)
  \Edge(w6)(v2)
  \Edge(v1)(w5)
 \Edge(v1)(w4)
  \Edge(v2)(w3)
   \tikzstyle{EdgeStyle}=[post,bend right,line width = 1]
\Edge(v1)(w3)
\Edge(v2)(w1)
 \Edge(v1)(w6)
 \Edge(v2)(w2)
\end{tikzpicture}
}
\end{center}
\caption{This image shows an $n/N_s$-central network with $n=2$ and $N_s=3$.  The graph consists of $8$ nodes denoted by $v_1,v_2, w_1, w_2, w_3, w_4, w_5, w_6$.  The set
$\{v_1, v_2 \}$ is the set of centers $\mathcal C$. The sets ${\mathcal S}({\mathcal C}), {\mathcal S}^*(v_1)$ and  ${\mathcal S}^*(v_2)$ are as follows:
${\mathcal S}({\mathcal C})=\{w_1,w_2, w_3, w_4, w_5, w_6\}$, ${\mathcal S}^*(v_1)=\{w_1, w_2, w_3\}$  and ${\mathcal S}^*(v_2)=\{w_4, w_5, w_6\}$.  The sets
${\mathcal S}^*(v_1) \cap  {\mathcal S}({\mathcal C}) =\{w_1, w_2, w_3\}$ and
${\mathcal S}^*(v_2) \cap  {\mathcal S}({\mathcal C}) = \{w_4, w_5, w_6\}$ contain three nodes each. \label{fig:1} }
\end{figure}
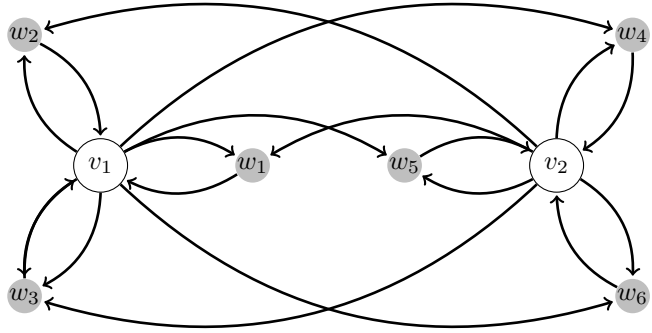

\section{Outline of main results}

Our results can be outlined as follows.
The result on the inertial dynamics existence describes a situation, when the interaction topology is quite arbitrary.  We assume that there exist $n$ slow nodes, say, $u_1,u_2,  ..., u_n$ with $\lambda_i=O(1)$ whereas
all the rest ones $u_{n+1}, ... u_N$ are fast, i.e.,  the corresponding $ \lambda_i$ have order $\Ord{\kappa^{-1}}$,   where $\kappa$ is a small parameter.  Then we show that there exists an inertial manifold  of dimension $n$.
   We obtain, under general conditions, that for times $t >> \kappa \log \, \kappa$ the dynamics of (\ref{Hop1})
is defined by the reduced equations
\begin{equation} \label{Hop2}
\frac{du_j}{dt}=F_j(u_1,... ,u_n,   W, h, \lambda), \quad
\end{equation}
\begin{equation} \label{Hop3}
u_k=U_k(u_1,... ,u_n,   W, h, \lambda), \quad k =n+1, ..., N,
\end{equation}
where $F_j$ and $U_k$ are some smooth functions of $u_1,..., u_n$, {and
$h,\lambda$ denote the vector parameters $(h_1,..., h_N)$ and $(\lambda_1,..., \lambda_N)$, respectively.} So, $F$ gives us
the inertial form on an inertial manifold. The inertial form completely defines the dynamics for large times \cite{Temam}.

More interestingly, we can show that the vector field  $F$ is, in a sense, maximally flexible. Roughly speaking,
by the number of nodes $N$ , the matrix  $W$  and $h$ we can obtain all possible fields $F$ (up to a small accuracy {$\epsilon$}, which can be done arbitrarily small as $N$ goes
to $ \infty$), see section \ref{flex} for a formal statement of this flexibility property.
For the networks  this flexibility property holds under {\bf $n/N_s$-Centrality assumption}.

Let us  introduce  a special control
parameter $\xi$, which modulates the
degradation  coefficient $\lambda_i$ for a hub: $\lambda_i=\xi \bar \lambda_i$ for some $i \in {\mathcal C}$.  This hub is a ''controller".
 When we vary the coefficient $\xi$,
 the interaction topology and the entries of the interaction matrix do not change,
but  the response time of the controller hub  changes.

One can choose the network parameters $N, W, \lambda$ in such a way that
for  $\xi>\xi_0$ the global attractor is trivial, it is a rest point, but for an open set of other values $\xi$  the global attractor of (\ref{Hop1}) contains  a number of local attractors. 

This result can be interpreted as ``maximal switchability''.
A similar effect was found in \cite{Deco} by numerical simulations for some models of neural networks.  This effect describes a transition from neural  resting states (NRS) to complicated global attractors, which occur as a reaction on learning tasks.
Note that in \cite{Deco} attractors consist of a number of  steady states.
In our case the global attractors can include {many local attractors of all possible kinds including chaotic and periodic ones}.

{We end this section with a remark. Our method approximates vector fields by neural networks, but what can be said about} the relationship between the trajectories of the simulated system and {the ones corresponding to the} neural network?

For chaotic and even for periodic attractors, direct comparison of trajectories is not a suitable
test for the accuracy of the approximation.
General mathematical arguments allow us say only that these trajectories will be close for bounded times.  For large times we can say nothing {especially for general chaotic attractors. } Consider the case when the
attractor $\mathcal A$ of the simulated system is transitive. This means the  dynamics is ergodic and for smooth function  $\phi$ the time averages
\begin{equation} \label{SFi}
S_{F, \phi}=\lim_{T \to +\infty} T^{-1} \int_{0}^T \phi({v}(t)) dt
\end{equation}
coincide  with the averages
$\int_{\mathcal A} \phi({v})  d\mu({v})$ over the attractor, where $\mu$ is an invariant measure {on $\mathcal A$}.

Then, a suitable criterion of approximation is
that the averages $S_{F, \phi}$ and the corresponding ones generated by the approximating centralized neural network, are close for smooth $\phi$:
\begin{equation} \label{approxNN}
|S_{F,  \phi} - S_{G_{anN}}, \phi|=Err_{approx} < \delta(\epsilon, \phi)
\end{equation}
where {$G_{anN}$ is the neural network approximation of $F$} and $\delta \to 0$ as $\epsilon \to 0$.
This ``stochastic stability'' property holds for hyperbolic (structurally stable) attractors \cite{kifer1986general,young1986stochastic, Viana}.

\section{Conditions on network parameters and attractor existence}

Our first results do not use any assumptions on the network topology. However, we suppose that
there are two types of network components that are distinguished by their time scales into
slow nodes and fast nodes.
 To take into account the two types of the nodes,
 we use distinct variables $v_j$  for slow variables, $j=1,\ldots,n$  and  $w_i$  for the fast ones, $i=1,\ldots, N-n=N_1$.
The real matrix entry
$A_{ji}$ defines the intensity of the action of
the fast node $i$
on the slow node $j$.  Similarly, the $n\times N_1$ matrix ${\bf B}$,  $N_1 \times N_1$ matrix $\bf C$ and $n \times n$ matrix ${\bf D}$
define the action  of the slow nodes on the fast ones, the interactions between the fast nodes   and the interactions between the slow nodes, respectively. We denote by $ h_i$ and $\lambda_i$
the  threshold and degradation parameters of the fast nodes and by $\tilde h_i$ and $\tilde \lambda_i$ the same parameters
for the slow nodes, respectively.
To simplify formulas, we use the notation
$$
   \sum_{j=1}^n D_{ij} v_j = {\bf D}_i v,  \quad
   \sum_{k=1}^N C_{jk} w_k={\bf C}_j w.
$$
Then,  equations (\ref{Hop1}) can be rewritten as follows:
\begin{equation}
\frac{dw_i}{dt} =
 \sigma\left(  {\bf B}_i v  + {\bf C}_i w - \tilde h_i\right) - \kappa^{-1} \tilde \lambda_i w_i,
\label{cn1}
\end{equation}
\begin{equation}
\frac{dv_j}{dt} =
\sigma \left({\bf A}_j w  + {\bf D}_j v -  h_j\right) -  \lambda_j v_j,
\label{cn2}
\end{equation}
where  $i=1,..., N_1, \ j=1,..., n$.
Here  unknown functions
 $w_i(t), v_j(t)$ are defined for times $t \ge 0$. We assume that   $\kappa$ is a positive parameter, therefore,  the variables $w_i$ are fast.

We set the initial conditions
 \begin{equation}
w_i(0)=\tilde \phi_i\ge 0, \quad v_j(0)= \phi_j\ge 0.
\label{initdat}
\end{equation}
 It is natural to assume that all concentrations are non-negative at the initial moment.
 It is clear that  they stay non-negative for all times.

\subsection{Global attractor exists}
\label{gla}

Let us prove that the network dynamics is correctly defined for all $t$ and solutions are non-negative and bounded.
For positive vectors $r=(r_1,..., r_n)$ and $R=(R_1, ...., R_{N_1})$,  let us introduce  the sets ${\mathcal B}$ defined by
$$
{\mathcal B}(r,  R)=\{(w, v): 0 \le v_j \le  r_j, \ 0 \le w_i \le R_j, \ j=1,...,n, \
 i=1, ..., N_1\}.
$$
Note that
$$
\frac{dw_i}{dt} < 1
 - \kappa^{-1} \tilde \lambda_i w_i.
$$
Thus, $w_i(t) < X(t)$ for positive times $t$, where
$$
\frac{dX}{dt} = 1
 - \kappa^{-1} \tilde \lambda_i X,     \quad X(0)=w_i(0).
$$
 Therefore, resolving the last equation, and repeating the same estimates for
$v_i(t)$, one finds
\begin{equation}
\begin{split}
 0 \le w_i(x, t) \le \tilde \phi_i \exp(-\tilde  \kappa^{-1}\lambda_i t) +  \kappa \tilde  \lambda_i^{-1}
 (1-\exp(- \kappa^{-1}\tilde \lambda_i t)), \\
 0  \le v_j(x, t) \le \phi_j \exp(-\lambda_j t) + \lambda_j^{-1}(1-\exp(-\lambda_j t)),
 \end{split}
\label{est0}
\end{equation}
Let us take arbitrary $a>1$ and let  $r_j(a)= a \lambda_j^{-1}$ and $R_i(a)=a \kappa \tilde \lambda_i^{-1}$.
Estimates (\ref{est0}) {show} that solutions of (\ref{cn1}), (\ref{cn2})  exist for all times $t$ and they enter
 the set ${\mathcal B}(r(a), R(a))$ at a time moment $t_0$. The solutions  stay in this set for all $t > t_0$, thus, this set is absorbing.  This shows that  system (\ref{cn1}),(\ref{cn2})
defines a global
dissipative semiflow $S_H^t$ \cite{Hale}.  Moreover,  this semiflow  has a global attractor contained in each ${\mathcal B}(r(a),R(a))$, where  $a>1$.

\subsection{Assumptions for slow/fast networks.}

A simpler asymptotic description of system dynamics is possible
 under  assumptions on network components timescales.
We suppose here that the $u$-variables are fast and the $v$-ones are slow.
We show then that the fast $w$ variables are slaved, for large times,
by the slow $v$ modes. More precisely, one has
 $w={ \kappa} U(v) + \tilde w$, where ${ \kappa} U(v)$ is a correction  and $ \kappa >0$ is a small parameter. This means
 that, for large times, the fast nodes dynamics
 is completely controlled  by the slow nodes.

To realize this approach,
let us assume that the system parameters ${\bf P}=\{{\bf A}, {\bf B}, {\bf C}, {\bf D}, h, \tilde h,\tilde  \lambda,  \lambda\}$ satisfy the following conditions:
\begin{equation}\label{cn3a}
{\bf  A}= \kappa^{-1} \bar {\bf  A},
\end{equation}
\begin{equation}
 \ | \bar {\bf A}|,  |{\bf B}|,  |{\bf C}|,  |{\bf D}| < c_0,
\label{cn3}
\end{equation}
\begin{equation}
 0 < c_1 < \bar\lambda_i   < c_2, \quad  0 < \tilde \lambda_i < c_3.
\label{cn4}
\end{equation}
Here  all positive constants $c_k$ are independent of $\kappa$ for small $\kappa$.

The scaling assumption on ${\bf A}$ is needed because, as we will prove later, $w = \Ord{\kappa}$ for small $\kappa$.
For the same reasons, ${\bf C}_i w$ can be neglected with respect to  ${\bf B}_i v$ for small $\kappa$, meaning that the
 action of  centers on  satellites is dominant with respect to satellites mutual interactions. In other words, these
conditions describe a divide and rule  control principle .

\section{Realization of prescribed dynamics and maximally flexible systems}\label{flex}

Our  goal is to show that the network dynamics
can realize, in a sense, arbitrary structurally
stable dynamics of the centers.
To precise this assertion,
let us describe the method of realization of the vector fields
for dissipative systems (proposed in  \cite{P1}).
More precisely, we are interested in
systems  enjoying
the following  properties:
\vspace{0.2cm}

{\bf A} {\sl These systems generate global
semiflows $ S_{\mathcal P}^t$
in an ambient Hilbert or Banach phase
space $H$. These semiflows depend on some
parameters $\mathcal P$ (which could be elements of another
Banach space $\mathcal B$).
They have global attractors
and finite dimensional local attracting invariant
$C^1$ - manifolds $\mathcal M$, at least for some $\mathcal P$}.

{\bf B} {\sl
Dynamics of
$S^t_{\mathcal P}$ reduced
on these
invariant manifolds can be, in a sense, almost completely tuned by variations of the parameter ${\mathcal P}$.

It can be described as follows. Assume the differential
equations
\begin{equation}
\label{2.1}
\frac{dq}{dt}=Q(q), \quad Q \in C^1({ B}^n)
\end{equation}
define a global semiflow in a unit ball
$B^n  \subset {\mathbb R}^n$.

For any prescribed dynamics (\ref{2.1}) and any $\epsilon >0$,
we can choose suitable parameters ${\mathcal P}={\mathcal
 P}(n, F, \epsilon)$ such that

{\bf B1} The semiflow $ S_{\mathcal P}^t$ has a $C^1$-
smooth locally attracting invariant manifold ${\mathcal M}_{\mathcal P}$
diffeomorphic to ${B}^n$;

{\bf B2}
The reduced dynamics $ S_{\mathcal P}^t\vert_{{\mathcal M}_{\mathcal P}}$
is defined by equations
\begin{equation}
\frac{dq}{dt}=\tilde Q(q, {\mathcal P}), \quad \tilde Q \in C^1( B^n)
\label{tQ}
\end{equation}
where the estimate
\begin{equation}
|Q -\tilde Q|_{C^1({ B}^n)} < \epsilon
\label{est1}
\end{equation}
holds. In other words, one can say that, by $\mathcal P$,
the reduced dynamics on the invariant manifold can be specified to
within an arbitrarily small error. }

Therefore, roughly speaking all robust dynamics (stable under small perturbations)
 can be generated by the systems, which satisfy above formulated properties.  Such systems can be named
{\em maximally flexible}. In order to {show that} maximal flexibility {covers also the case of chaotic dynamics}, let us recall some facts about chaos and hyperbolic sets.

Let us consider dynamical systems  (global semiflows) $S_1^t, ..., S_k^t$, $t>0$,  defined on the $n$-dimensional closed ball $B^n \subset {\mathbb R}^n$ defined by
     finite dimensional vector fields $F^{(k)} \in C^1(B^n)$ and having  structurally stable attractors ${\mathcal A}_l$, $l=1,..., k$.  These attractors
can have a complex form, since it is well known
that structurally stable dynamics may be ``chaotic".
There is a rather wide variation in different definitions of
 "chaos". In principle, one can use here
any concept of chaos, provided that this is stable
under small $C^1$ -perturbations. To fix ideas, we shall use here,
following
\cite{Ruelle}, such a definition. We say that a finite
dimensional dynamics is chaotic if it
generates a compact invariant hyperbolic set $\Gamma$, which is not a  periodic cycle or a rest point (for a definition of hyperbolic
sets see, for example, \cite{Ruelle}).
 The hyperbolic sets give remarkable analytically tractable examples, where chaotic dynamics can be studied. For example, the Smale horseshoe  is a hyperbolic set.
If
this set $\Gamma$ is attracting we say that $\Gamma$
is a chaotic (strange) attractor.
In this paper,  we use only the following
basic property of hyperbolic sets,
  so-called Persistence \cite{Ruelle}.
This means that the hyperbolic sets are, in a sense, stable(robust).  This property can be described as follows. Let  a system of differential equations be defined by a  $C^1$-smooth vector field $Q$
on an open domain in ${\mathbb R}^n$ with a smooth boundary or on a smooth compact finite dimensional manifold.
Assume this system  defines a dynamics having a compact invariant hyperbolic set $\Gamma$. Let us consider $\epsilon$-perturbed  the vector field $Q + \epsilon \tilde Q$
, where $\tilde Q$ is bounded in $C^1$-norm. Then, if $\epsilon>0$ is sufficiently small, the perturbed field also generates  dynamics with another compact invariant
hyperbolic set $\tilde \Gamma$. The corresponding dynamics restricted to $\Gamma$
and $\tilde \Gamma$ respectively, are topologically orbitally equivalent (  topological equivalency of two semiflows means that there exists a homeomorphism, which
maps  the  trajectories of  the first semiflows on the trajectories of the second one, see \cite{Ruelle} for details).


We recall that chaotic structurally stable  ( persistent) attractors and invariant sets exist:
this fact  is well known from the theory of hyperbolic dynamics \cite{Ruelle}.

   Thus, any kind of the chaotic hyperbolic sets
 can occur in the dynamics of the systems, for example,
the Smale horseshoes, Anosov
flows, and the Ruelle-Takens-Newhouse chaos, see \cite{Ruelle}.
Examples of systems satisfying these properties can be presented
 by some  reaction-diffusion equations
and systems  \cite{ P1, Vak1, Vak2},
and  neural network models   \cite{Vak2}.

\section{Main results}

For  vectors $a=(a_1,  ..., a_n)$ and $b=(b_1,..., b_n)$ such that $a_i < b_i$ for each $i$  let us denote by
\begin{equation} \label{piab}
\Pi(a, b)=\{v \in {\mathbb R}^n:   a_i \le v_i \le b_i \}
\end{equation}
 a $n$-dimensional box in $v$-space.
Moreover, let us define   $\Pi_{\lambda}$ by $\Pi_{\lambda}=\Pi(0, \lambda^{-1}) $, where the vector $\lambda^{-1}$ has  components
$(\lambda_1^{-1}, ..., \lambda_n^{-1})$.

\begin{theorem} \label{T1}
{ Under assumptions  (\ref{eq2.5}),  (\ref{cn3a}), (\ref{cn3})  and (\ref{cn4})  for sufficiently small  $\kappa$ there exists a  $n$-dimensional inertial manifold  ${\mathcal M}_n$ defined by
\begin{equation}
 w_i=\kappa {\tilde \lambda}_i^{-1} U_i(v, \kappa, {\bf P}) , \quad  v     \in \Pi_{ \lambda}
\label{inert}
\end{equation}
where $U_i \in C^{1+r}(\Pi_{ \lambda})$, and $ r \in (0,1)$.  The functions $U_i$ admit the estimate
\begin{equation}
|U_i(v, \kappa, {\bf P}) - \sigma\left(  {\bf B}_i v  - \tilde h_i\right)|_{C^1(\Pi_{ \lambda})} <  c_4 \kappa,  \quad  v     \in \Pi_{ \lambda}.
\label{inert2}
\end{equation}

The $v$ dynamics for large times  takes the form
\begin{equation}
\frac{dv_j}{dt} = F_j( v,  {\bf P}) + \tilde F_j(v, \kappa, {\bf P}),
\label{reddyn}
\end{equation}
 where $\tilde F_j$ satisfy
\begin{equation}
\label{esttildeF}
|\tilde F_j|_{C^1(\Pi_{\lambda})}  < c_6\kappa
\end{equation}
with
\begin{equation}  \label{Fj}
F_j( v, {\bf P}) =
 \sigma\left(
\sum_{i=1}^{N-n} \bar { A}_{ji}{\tilde \lambda_i}^{-1} \sigma\left(  {\bf B}_i v  - \tilde h_i\right)  + {\bf D}_j v  - h_j \right) -  \lambda_j v_j.
\end{equation}
}
\end{theorem}

Note that the matrix ${\bf C}$ is not involved in  relation (\ref{Fj}), which defines the family of the vector fields  $F$  (  inertial forms). This property holds due to
 the property that  inter-satellite interactions are dominated by the satellite-center ones.
The next assertion means that this principle allows us to create a   network dynamics with prescribed dynamics (if the network  satisfies $n/N_s$-centrality assumption and
$N$ is large enough).
 It  is valid under the additional condition that the interaction graph $(V, E)$ verifies the centrality condition.

\begin{theorem} \label{T2} Assume $n/N_s$-centrality assumption is satisfied. Then the family of the vector fields $F$ defined by (\ref{Fj})  is dense in the set of
  all  $C^1$  vector fields $Q$ defined on the unit  ball $B^n\subset {{\mathbb R}}^n$.  In the other words,  centralized Hopfield neural networks
are maximally flexible.
\end{theorem}

Let us choose some $i_C$ such that $i_C$ belongs to  ${\mathcal C }$. The corresponding node
 will be called a controller hub.  We introduce the control parameter  $\xi$ by
\begin{equation}
\lambda_{i_C}=\xi \bar \lambda_{i_C},
\end{equation}
where we fix a positive $\bar \lambda_{i_C}$.

Theorem \ref{T2} can be used to show the following

  \begin{theorem}  \label{T3}  {\bf (Maximal switchability theorem)}   {Let us consider   dynamical systems  (global semiflows) $S_1^t, ..., S_k^t$, $t>0$,  defined on the $n$-dimensional closed ball $B^n \subset {{\mathbb R}}^n$ defined by
     finite dimensional vector fields $F^{(k)} \in C^1(B^n)$ and having  structurally stable attractors ${\mathcal A}_l$, $l=1,..., k$.

For sufficiently large $N$ and  any graph $(V, E)$ satisfying the $n/N_s$- centrality condition there  exists a choice of  interactions $W_{ij}$  and thresholds $h_i$ such that
Assumption 1 holds and

({\bf i})
there exist a $\xi_0$ such that for all   $\xi > \xi_0$ the dynamics of network (\ref{Hop1}) has a rest point, which is a global attractor;

({\bf ii})
  for an open interval of values  $\xi$  the global  semiflow $S_H^t$ defined by (\ref{Hop1})
have   local  attractors $\mathcal B_l$ such that the restrictions of
 the   semiflow  $S_H^t$ to $\mathcal B_l$  are orbitally topological equivalent to the semiflows  $S^t_l$ restricted to   $ \mathcal A_l$}.
\end{theorem}

Finally, let us give an estimate on  the maximal number of equilibria $N_{eq}$ of centralized networks. This number is a characteristics of  the network capacity, flexibility and
adaptivity.
To proceed to these estimates, let us define a procedure, which can be named
decomposition into ``distar'' motifs.
 In the network interaction graph $(E, V)$ we choose some nodes $v_1,..., v_n$, which we conditionally consider as hubs. By ``distar'' motif we understand a part of interaction graph consisting of
the hub $v_j$ and the subset $S_j$ of the set $S^*_j$ (defined by (\ref{centrality})) consisting of the nodes connected
in both directions to $v_j$:
$S_j=\{ v_i  \in V :  \  (i,j) \ and \ (j, i) \in E \}$.
  This distar motif becomes an usual star if directions of the edges are ignored.
Consider the union $U_n$ of all $S_j$. Some nodes $w \in U_n$ may belong to two different sets $S_j$ and  $S_k$, where $k \ne j$.
    We remove from the vertex set $V$ all such nodes.
After such removing  we obtain a part of graph $G_n=(V', E')$ of the initial graph $(E, V)$, which is a union of $n$ disjoint distars $S_1,..., S_n$,
where each $S_k$ contains a single center $\{v_k\}$ and $\mu(S_k)$ satellites connected with the center in both directions.  Recall that
the graph $(V', E')$ is a part of graph $(V, E)$ if $V'\subset V$ and $E' \subset E$.
These numbers $\mu(S_k)$ depend on the choice of hub nodes $\{v_1, ..., v_n\}$.

We will prove the following theorem:

 \begin{theorem}  \label{TNeq}
{The maximal possible number $N_{eq}(E, N)$ of equilibria of a network with a given  interaction graph $(E, V)$, where $V$ consists of $N$ nodes,
satisfies
\begin{equation} \label{NeqFS}
 N_{eq} \ge \sup  \mu (S_1) \mu(S_2) ... \mu(S_n),
\end{equation}
 where the supremum is taken over all integers $n >0$ and all  graphs $G_n$,which are parts of interaction graph $(V, E)$  and consist of $n$ disjoint distars.  Here $\mu(S_l)$ is the number of the nodes in the  distar  $S_l$.
}
\end{theorem}

Consider now  graphs, which are unions of identical distars. The degree of the center of each  distar
is $\floor{(N-n)/n}$.
Then, the  maximal possible number $N_{eq}$ of equilibria in such a centralized network (\ref{Hop1}) with $N$ {nodes} and $n$ centers
satisfies
$
 N_{eq} \ge  \floor{(N-n)/n}^n,
$
where  $\floor{x}$  denotes the floor of a real number  $x$.
Note that for a fixed $N$ the maximum of $(N/n)^n$ over $n=1,2,...$ is attained at $n=\floor{N/5}$, when the {distars contain  $5$ satellites each}. Therefore we obtain the
estimate  $N_{eq} \ge 4^{\floor{N/5}}$.

\section{Proof of Theorem \ref{T1}}

Let us start by proving a lemma
\begin{lemma}
{
Under assumptions (\ref{cn3a}), (\ref{cn3}) and  (\ref{cn4})
 for sufficiently small positive $\kappa < \kappa_0$ solutions $(u, v)$ of (\ref{cn1}), (\ref{cn2})
  and (\ref{initdat})
  satisfy
\begin{equation}
w_i(t)=\kappa U_i(v(t), {\bf B}, \tilde h) + \tilde w_i(t),
\label{cn5}
\end{equation}
where
 $U=(U_1, ..., U_n)$ is defined by
\begin{equation}
  U_i(v, {\bf B}, \tilde h)= \tilde \lambda_i^{-1}  \sigma\left({\bf B}_{i} v( t)  - \tilde h_i \right).
\label{cn6}
\end{equation}
Then, for some $T_0$ function $\tilde w$ satisfies the estimates
\begin{equation}
|\tilde w(t)| < c_1\kappa^2,  \quad  t > T_0
\label{cn8}
\end{equation}
where $c_1$ does not depend on $t$ and $\kappa$.  The time moment $T_0$ depends on initial data and the network parameters.
}
\end{lemma}

{\bf Proof}. Let us introduce a new variables
$\tilde w_i$ by (\ref{cn5}). They satisfy the equations
\begin{equation}  \label{utilde}
\frac{d\tilde w_i}{dt}= H_i(v, \tilde w) - \kappa^{-1} \tilde \lambda_i  \tilde w_i,
\end{equation}
where
$$
H_i(v, \tilde w)=\kappa Z_i (v)  + W_i (v, \tilde w),
$$
$$
Z_i (v) =\sum_{j=1}^n \frac{\partial U_i(v)}{\partial v_j}(
\sigma \left(\bar {\bf A}_j U  + {\bf D}_j v -  h_j\right) - \xi \bar  \lambda_j v_j) ,
$$
and
$$
W_i (v, \tilde w) =  \sigma\left(  {\bf B}_i v  + {\bf C}_i w - \tilde h_i\right) - \sigma\left(  {\bf B}_i v  - \tilde h_i\right).
$$
Let us estimate $H_i(v, \tilde w)$ for sufficiently large $t$. According to (\ref{est0}), for such times we can
use that
$(w,v) \in {\mathcal B}( r(a), R(a))$, where $a >1$.  In this domain ${\mathcal B}( r(a), R(a))$ one has $\sup |Z_i| < c_2$
and $\sup|W_i| < c_3\kappa$, where $c_2, c_3$ are independent of $\kappa$.  Therefore,
$$
H_i(v(t), \tilde w(t)) < c_0  \kappa,  \quad t  > T_0(\kappa, {\bf P}).
$$
Now, as above in subsection  \ref{gla},  equation (\ref{utilde}) entails  estimate   (\ref{cn8}).  The assertion is proved.

{\bf Proof of Theorem 6.1.}
The rest part of the proof
 of Theorem \ref{T1}  uses the well known technique
of invariant manifold theory, see, for example, \cite{Ruelle, Temam,  He}.
Let us consider the domain  $D_{\kappa}=\{w: |w| < c_1 \kappa^2$\}. Theorem 6.1.7 \cite{He} shows  that  for $d \in (0,1)$
 there is a locally attractive $C^{1+d}$- smooth invariant manifold ${\mathcal M}_n$.
  Relation (\ref{inert2})  follows from (\ref{cn8}). The global attractivity of this manifold also follows from (\ref{cn8}).
The theorem is proved.

\section {Proof of Theorems \ref{T2},   \ref{T3} and \ref{TNeq}}

\subsection{Proof of  Theorem \ref{T2}}

The main idea of the subsequent statement is to study the dependence of the fields
$F_j$ defined by Eq.\eqref{Fj} on the parameters ${\bf P}$.  To this end, we apply a special method stated in the next subsection.

Let us formulate  a lemma, that gives us a key tool and which implies Theorem \ref{T2}.

\begin{lemma} \label{mainlemma}
{ Assume
\begin{equation}
a_i >  \delta/\lambda_i,  \quad b_i <  (1-\delta)/\lambda_i \quad i=1, ...,n.
\label{ab}
\end{equation}

 Let $Q=(Q_1(v), ..., Q_n(v))$ be a $C^1$ smooth vector field
on $\Pi(a, b)$ and $\delta >0$ verify
\begin{equation} \label{estim0}
 -\delta < Q_i(v) < \delta,\quad v \in \Pi(a,b), \quad i=1,..., n.
\end{equation}

 Then there are  parameters
${\bf P}$ of the neural network
such that the field $F$ defined by  (\ref{Fj}) satisfies the estimates
\begin{equation} \label{estim1}
\sup_{v \in \Pi(a, b)} |F(v, {\bf P}) -  Q(v)| < \epsilon,
\end{equation}
\begin{equation} \label{estim2}
\sup_{v \in \Pi(a, b)} |\nabla F(v, {\bf P}) - \nabla Q(v)| < \epsilon.
\end{equation}
In other words, the fields $F$ are dense in the vector space of all $C^1$ smooth vector fields satisfying
to (\ref{estim0}).
}
\end{lemma}

{\bf Proof}. The proof uses the standard results of the multilayered network theory.

{\em Step 1}.
The first preliminary step is as follows. Let us solve the system of equations
\begin{equation} \label{fund}
\sigma(R_j) = Q_j(v)+\lambda_j v_j,  \quad v \in \Pi(a, b)
\end{equation}
with unknown $R_j$. Here $R_j$ are the regulatory inputs of the sigmoidal functions.
These equations have a unique solution
due to conditions (\ref{eq2.5}), (\ref{ab}) and (\ref{estim0}): the right hand sides
$V_j +\lambda_j v_j$ range in $(0,1)$. The solutions $R_i(v)$ are $C^1$-smooth vector fields.

{\em Step 2}.
Consider relation (\ref{Fj}).
We choose entries $A_{ji}$ and $B_{il}$ in a special way. First, let us set
$A_{ji}=0$ if $ i \notin  {\mathcal S}^*(j)$, where the set $ {\mathcal S}^*(j) $ is defined in the $n/N_s$-centrality assumption,  see condition {\bf ii}. Recall that  $ {\mathcal S}^*(j) $
is the set of the satellites acting on the center $j$.
Note that then
 sum (\ref{Fj}) can be rewritten as
\begin{equation}  \label{Fj2}
F_j( v, {\bf P}) =
 \sigma\left(
\sum_{i \in {\mathcal S}^*(j)} \bar { A}_{ji}{\tilde \lambda_i}^{-1} \sigma\left(  {\bf B}_i v  - \tilde h_ i\right)  + {\bf D}_j v  - h_j \right) -  \lambda_j v_j.
\end{equation}

Using the result of step 1 and this relation, we see that our problem is reduced to the following:
to approximate $R_j(v)$ in $C^1$ norm with a small accuracy $O(\epsilon)$ by
\begin{equation}  \label{Hj2}
H_j( v, {\bf P}) =
\sum_{i \in {\mathcal S}^*(j)} \bar { A}_{ji}{\tilde \lambda_i}^{-1} \sigma\left(  {\bf B}_i v  - \tilde h_i\right)  + {\bf D}_j v  - h_j.
\end{equation}
Note that, according to
the centrality assumption,  the set   ${\mathcal S}^*(j)$  contains $N_s > CN^{\theta}$ elements.  Moreover, due to this  assumption, the sum ${\bf B}_i=\sum_k B_{ik} v_k$
involves all $k, \ k=1,..., n$.
Therefore, since $n$ is fixed and $N$ can be taken arbitrarily large, the theorem on the universal approximation by multilayered perceptrons (see, for example, \cite{Barron}) implies that
the fields
$H=(H_1, ..., H_n)$
are dense in the Banach space of all the vector fields on $\Pi(a,b)$ (with $C^1$- norm).
Therefore, $H_j$ approximate $R_j$ with  $O(\epsilon)$-accuracy in $C^1$- norm. This finishes the proof.

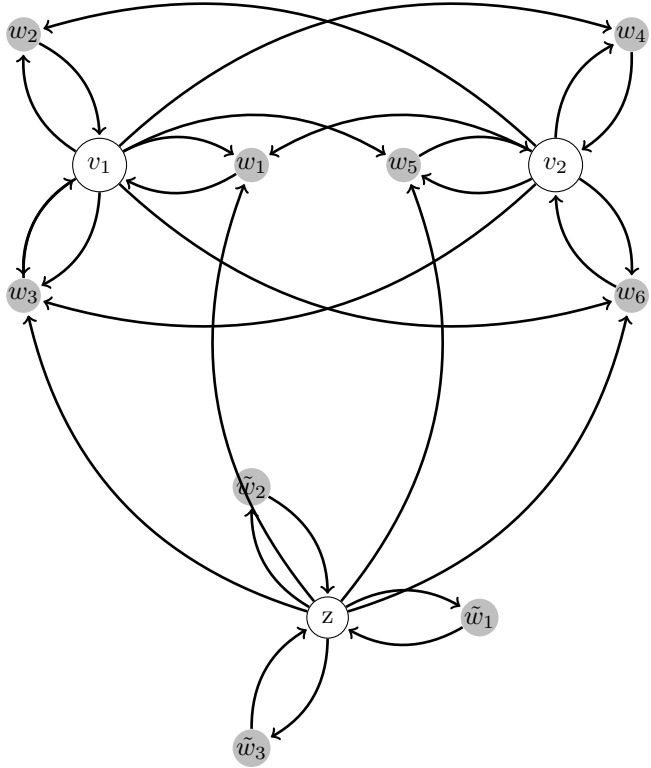
\begin{figure}[h!]
\begin{center}
\scalebox{1}{
\begin{tikzpicture}
 \SetUpEdge[lw         = 0.5pt,
            color      = black,
            labelstyle = {fill=white, sloped}]
  \tikzset{node distance = 2.5cm,main node/.style={circle,fill=blue!12,draw,font=\sffamily\large\bfseries}
   }
 \tikzstyle{vertex}=[circle,fill=black!25,minimum size=7pt,inner sep=0pt]
  \node[vertex, circle,xshift=-3cm] (w1) at ({360/3 * 0 }:2) {$w_1$};
 \node[vertex, circle,xshift=-3cm] (w2) at ({360/3 * 1 }:2) {$w_2$};
 \node[vertex, circle,xshift=-3cm] (w3) at ({360/3 * 2 }:2) {$w_3$};
 \node[draw, circle,xshift=-3cm] (v1) at (0:0) {$v_1$};
 \node[vertex, circle,xshift=3cm] (w4) at ({360/3 * 0+ 60}:2) {$w_4$};
 \node[vertex, circle,xshift=3cm] (w5) at ({360/3 * 1+ 60}:2) {$w_5$};
 \node[vertex, circle,xshift=3cm] (w6) at ({360/3 * 2+ 60}:2) {$w_6$};
 \node[draw, circle,xshift=3cm] (v2) at (0:0) {$v_2$};
   \node[vertex, circle,yshift=-6cm] (w7) at ({360/3 * 0 }:2) {$\tilde w_{1}$};
 \node[vertex, circle,yshift=-6cm] (w8) at ({360/3 * 1 }:2) {$\tilde w_{2}$};
 \node[vertex, circle,yshift=-6cm] (w9) at ({360/3 * 2 }:2) {$\tilde w_{3}$};
 \node[draw, circle,yshift=-6cm] (z) at (0:0) {z};

 \tikzstyle{EdgeStyle}=[post,bend left,line width = 1]
 \Edge(v1)(w1)
  \Edge(w1)(v1)
   \Edge(v1)(w2)
  \Edge(w2)(v1)
   \Edge(v1)(w3)
  \Edge(w3)(v1)
   \Edge(v2)(w4)
  \Edge(w4)(v2)
   \Edge(v2)(w5)
  \Edge(w5)(v2)
   \Edge(v2)(w6)
  \Edge(w6)(v2)
  \Edge(v1)(w5)
 \Edge(v1)(w4)
  \Edge(v2)(w3)
   \Edge(z)(w7)
 \Edge(z)(w8)
  \Edge(z)(w9)
      \Edge(w7)(z)
    \Edge(w8)(z)
    \Edge(w9)(z)
 \Edge(z)(w1)
 \Edge(z)(w3)
   \tikzstyle{EdgeStyle}=[post,bend right,line width = 1]
\Edge(v1)(w3)
\Edge(v2)(w1)
 \Edge(v1)(w6)
 \Edge(v2)(w2)

  \Edge(z)(w5)
 \Edge(z)(w6)

\end{tikzpicture}
}
\end{center}
\caption{Modular architecture. The switching module consists of the center $z$ and the satellites {$\tilde w_1, \tilde w_2, \tilde w_3$}. The generating module consists of the centers $v_1, v_2$ and the satellites $w_1,..., w_6$.}
\label{fig:2}
\end{figure}

\subsection{Proof of Theorem  \ref{T3}}

{\em Ideas behind proof}.
 Before stating a formal proof, we present a brief outline, which describes main ideas of the proof and the architecture of the switchable network.
The network consists of two modules.  The first module is a generating one and it is  a centralized neural network with $n$ centers $v_1,..., v_n$ and satellites $w_1, ...,w_N$.
The second module  consists of a center $v_{n+1}=z$ and $m$ satellites $\tilde w_1,...,  \tilde w_m$. The satellites from this module  interact only with the module center  $z$, i.e., in this module the interactions
can be described by a distar graph. Only the center of the second module interacts with the neurons of the first (generating) module. We refer {to} the second module as a switching one.  {This architecture is shown on Fig. \ref{fig:2}}.

For the switching module the correspoding equations have the following form.
Let us consider a distar interaction motif,  where a node $z$ is connected in both directions with $m$ nodes $\tilde w_1, ...,  \tilde w_m$.
We set $n=1$ and $N_1=m$,  $ \tilde \lambda_i =1$, ${\bf D}={\bf 0}$, ${\bf C}={\bf 0}$, $ \lambda_1=1$,  and $A_{1j}=\kappa^{-1} \bar a_j$
 in eqs.  (\ref{cn1}) and (\ref{cn2}).  By such notation the equations for the switching module can be rewritten in   the form
\begin{equation}
\frac{d\tilde w_i}{dt} =
 \sigma\left(\tilde b_i z  - \tilde h_i\right) - \kappa^{-1}  \tilde w_i,
\label{cn1S}
\end{equation}
\begin{equation}
\frac{dz}{dt} =
\sigma \left(  \kappa^{-1}\sum_{j=1}^m \tilde  a_j \tilde w_j   -  h\right) -   \xi \bar \lambda z,
\label{cn2S}
\end{equation}
where  $i=1,..., m$ and $\tilde b_i, \tilde   a_j, \bar \lambda >0$.

 Under above assumptions on the  network interactions, equations  for generating module can be represented as follows:
\begin{equation}
\frac{dw_i}{dt} =
 \sigma\left(  {\bf B}_i v  + {\bf C}_i w - d_i z - \bar h_i\right) - \kappa^{-1} \tilde \lambda_i w_i,
\label{cn1mo}
\end{equation}
\begin{equation}
\frac{dv_j}{dt} =
\sigma \left({\bf A}_j w  + {\bf D}_j v -  \tilde  d_j z - h_j\right) -  \lambda_j v_j,
\label{cn2mo}
\end{equation}
where  $i=1,..., N, \ j=1,..., m$ and $d_i, \tilde  d_j$ are coefficients.

These equations involve $z$ as a parameter. This fact can be used in such a way. Consider the system of the differential equations
\begin{equation} \label{QVV5}
 dv/dt=Q(v, z), \quad v = (v_1,\ldots,v_n)
\end{equation}
where $z$ is a real control parameter.  Let $z_1, ..., z_{m+1}$ be some values of this parameter.  We find a vector field $Q$ such that for $z=z_l$, where $l=1,...,m$,
the dynamics  defined by  \eqref{QVV5}  has the prescribed structurally stable invariant sets $\Gamma_l$. Furthermore, according to theorem \ref{T2}, for each positive $\epsilon$ we can choose the parameters $N, {\bf B}_i ,{\bf C}_i, \tilde b_i, \tilde  a_i, \bar h_i,  {\bf A}_j ,{\bf D}_j,d_i,  \tilde d_j, h_j, \lambda_j,\tilde \lambda_i$ of the system (\ref{cn1mo}) and (\ref{cn2mo}) such that the dynamics of this system will have structurally stable invariant sets $\tilde \Gamma_l$ topologically equivalent to $\Gamma_l$.

For the switching module we adjust the center-satellite interactions and the center response time parameter $\xi$ in such a way that for a set of values  $\xi$ the switching module has the dynamics of system  (\ref{cn1S}),(\ref{cn2S})
with $m$ different stable hyperbolic equilibria $z=z_1, z_2,..., z_{m+1}$  and   for sufficiently large $\xi $ system  (\ref{cn1S}) and (\ref{cn2S}) has  a single equilibrium close to $z_1=0$.  Existence of such a choice will be shown in coming lemma \ref{Flower}.
Then the both modules form a network having need dynamical properties formulated in the assertion of Theorem \ref{T3}.

{\bf Proof}. Let us formulate some auxiliary assertions.
First we consider the switching module.

\begin{lemma}\label{Flower}
{ Let $m$ be a positive integer and $\beta \in (0,1)$. For sufficiently small $\kappa >0$ there exist
$\bar a_j, b_i, \tilde h_i, h$ such that

{\bf i} for an open interval of values $\xi$
system (\ref{cn1S}),(\ref{cn2S})
 has $m$ stable hyperbolic rest points  $z_j \in (j -1+\beta, j +\beta)
$, where $j=1,..., m$;

{\bf ii} for $\xi > \xi_0 >0$ system (\ref{cn1S}),(\ref{cn2S})
 has a single stable hyperbolic rest point.
}
\end{lemma}

{\bf Proof}.
Let  $h=0$. To find equilibria $z$, we set $d\tilde w_i/dt=0$, and express $\tilde w_i$ via $z$. Then we obtain the following equation for the rest points $z$:
\begin{equation}
   \xi z=\sigma \left(  \sum_{j=1}^m \tilde a_j    \sigma\left(\tilde b_j z   - \tilde h_j\right)  \right).
\label{cn2SE}
\end{equation}
For especially adjusted parameters  eq. (\ref{cn2SE}) has at least $m$ solutions, which give
stable equilibria of system (\ref{cn1S}),(\ref{cn2S}). To show it,  we assume that  $0 < \kappa << 1$,   $\tilde  b_j =\tilde  b =\kappa^{-1/2}$
  and  $\tilde h_j= \tilde  b \mu_j$, where $\mu_j= j-1 +\beta$.  We obtain then
\begin{equation} \label{starway}
 V(\xi z)= \sum_{j=1}^m \sigma(\tilde  b(z -  \mu_j)) + O(\kappa)=F_m(z, \beta, \kappa),
\end{equation}
where $V(z)$ is a function inverse to $\sigma(z)$ defined on $(0,1)$.  Since $\tilde  b >> 1$ for small $\kappa$, the plot of the function
$F_m$ is close to a stairway (see  Fig. 3).   Let
$$
\xi=1, \quad  \tilde a_1=V(\mu_1)+\kappa, \quad   \tilde a_j=V(\mu_j) - V(\mu_{j-1}), j=2,...,m.
$$
The intersections of the curve $V (z)$ with the almost horizontal pieces of the plot of $F_m$ give us $m$ stable equilibria
of system (\ref{cn1S}),(\ref{cn2S}).  These equilibria $z_j$ lie in the corresponding intervals $(j -1+\beta, j +\beta)$.
  For sufficiently large $\xi$ we have a single rest stable point $z$ at $0$. The lemma is proved.
\vspace{0.2cm}

Consider   compact invariant hyperbolic sets $\Gamma_1, ...,  \Gamma_m$  of semiflows defined by arbitrarily chosen
$C^1$ smooth vector fields   $Q^{(l)}$ on the unit ball $B^n \subset {{\mathbb R}}^n$, where $l=1,..., m$.

\begin{lemma} \label{L3} {Let $\Pi(a, b)$ be a box in ${\mathbb R}^n$ and $m>1$ be a positive integer. There is a $C^1$-smooth vector field $Q$ on  $\Pi(a, b) \times [0, m+1]$ such that
equation (\ref{2.1}) defines a semiflow having hyperbolic sets $\Gamma_1, ...,  \Gamma_{m}$ and the restriction of this field on  $\Pi(a, b) \times [0, 1]$ has  an attractor consisting of a single hyperbolic
rest point}.
\end{lemma}

\hskip-0.75truecm

\begin{figure}[h!]
\centerline{
\includegraphics[width=80mm]{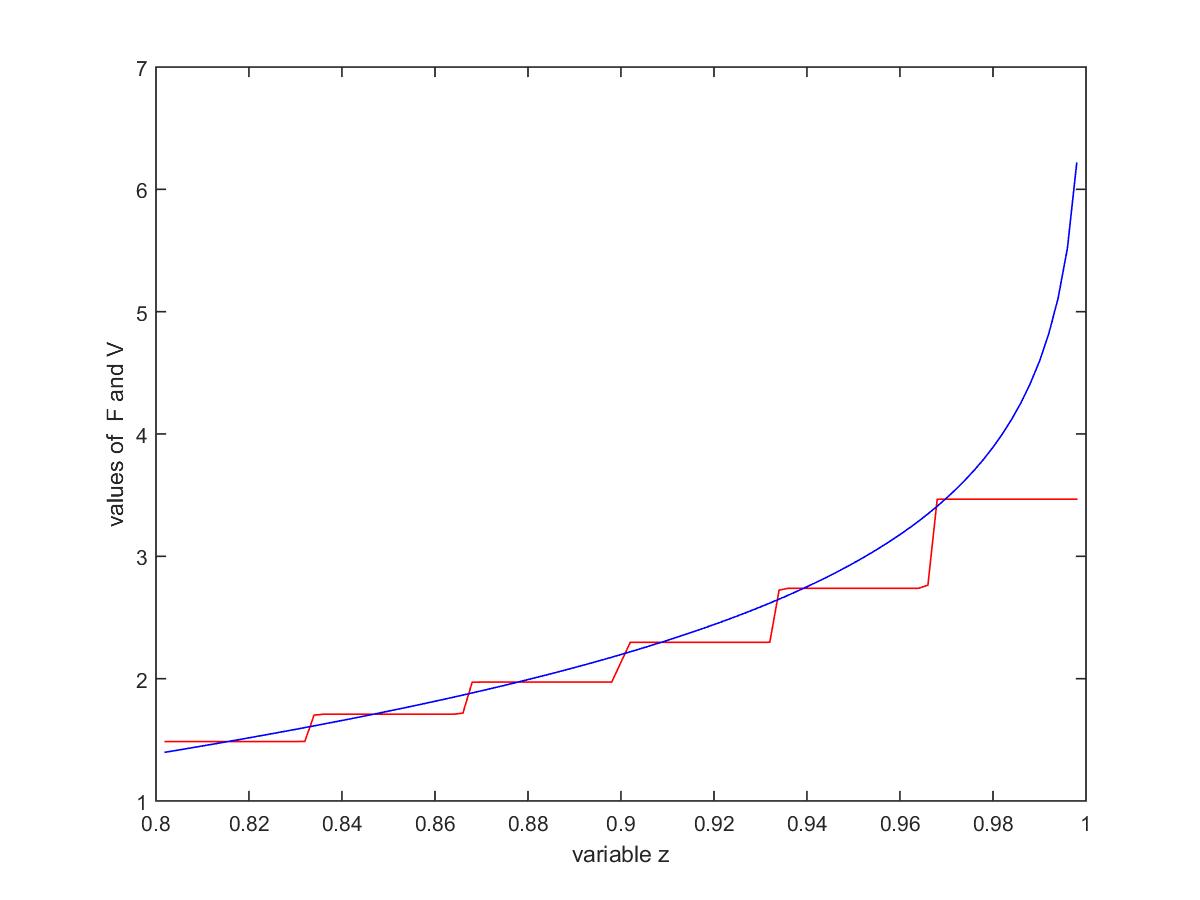}}
\caption{\small
The intersections of the curve $F_m(z, \beta, \kappa)$ and the curve $V(z)$ give equilibria of system \eqref{cn1S},\eqref{cn2S} for $\xi=1$. Stable equilibria
correspond to the intersections of $V$ with almost horizontal pieces of the graph of $F_m$. }
\label{fig:3}
\end{figure}

{\bf Proof}. The proof uses the following idea. For $k \in \{2, ..., m+1\}$  let $Q^{(k)}(v)$  be a vector field on $\Pi(a,b)$ having $\Gamma_{k-1}$ as an invariant compact hyperbolic set. Moreover, suppose that
$Q^{(1)}$ has a single globally attracting rest point in $\Pi(a,b)$,  $z_j \in (j-1+\beta, j+\beta)$, where $j=1,..., m$ and $\beta \in (0,1)$.
Let $\chi_{k}(z)$ be smooth functions of $z \in {{\mathbb R}}$ such that
$$
 \chi_{k}(z_l)=\delta_{lk}, \ l  \in \{1, ..., m\},  \quad k=1,..., m
$$
where $\delta_{lk}$ stands for the Kronecker delta.
Let $Q(v,z)$ be the vector field on $\Pi(a, b) \times [0, m+\beta]$  defined by
\begin{equation} \label{qvz}
 Q_i(v, z)=\sum_{k=1}^m Q_i^{(k)}\chi_{k}(z),  \quad i  \in \{1, ..., n\},
\end{equation}
for first $n$ components
and $n+1$-th component of this field (denoted by $z$) is defined by
\begin{equation} \label{qvz1}
 Q_{n+1}(v, z)=F_{m}(z, \beta, \kappa),
\end{equation}
where $F_{m}$ is defined by
(\ref{starway}).  For $\beta \in (0,1)$  the function $F_m$ has stable roots at the points $z=1,2,..., m$.
We observe that the equation for $z$-component
$dz/dt=F_{m}(z, \beta, \kappa)$
does not involve $v$. By applying Lemma \ref{Flower}
we note that  solutions $z(t,z(0))$ of the Cauchy problem for this differential equation verify
$
|z(t) - z_j| <  \exp(-c_1 t),
$
if $z(0)$ lies in an open  neighbourhood of $z_j$.
To conclude the proof, we consider the system
$$
 dv_i/dt=Q_i(v, z), \ i=1,..., n, \
$$
$$
dz/dt=F_{m}(z, \beta, \kappa) -\xi \bar \lambda z=Q_{n+1}(z).
$$
The right hand sides of this system define the field $Q$ of dimension $n+1$ from the assertion of Lemma \ref{L3}.
To check this fact, we apply Lemmas \ref{mainlemma} and \ref{Flower}  that completes the proof.
\vspace{0.2cm}

Next, to finish the proof of Theorem \ref{T3}, let us take a box $\Pi(a, b)$, where  $0 < a_i < b_i$. The semiflows defined by differential equations $dv/dt= \delta Q(v)$
are orbitally topologically equivalent for  all $\delta >0$.  We approximate the first $n$ components of the field $Q$ by our neural network using  Lemma \ref{L3} .
We multiply here $Q$  on an appropriate positive $\delta$ to have a field with components
bounded by sufficiently small number in order to apply Lemma \ref{mainlemma}.
Namely, we take $\delta$ such that
$a_i > \delta/(\xi_0  \bar \lambda_i)$ and $b_i < (1- \delta)/(\xi_1 \bar \lambda_i)$ and apply Lemma \ref{mainlemma}.
Note that this appoximation does not involve the control parameter $\xi$.  Indeed,  this parameter is involved only in the approximation of
$ Q_{n+1}$, which can be done independently, see  the distar graph lemma \ref{Flower}. This concludes the proof of  Theorem \ref{T3}.

{\em Remark}.  In Theorem \ref{T3}, we assume  that the vector field $Q(v)$ is given.  However, by centralized networks we can solve the problem of identification of dynamical systems
supposing that the trajectories $v(t)$ are given on a sufficiently large time interval whereas $Q$ is unknown or we know this field only up to unknown parameters.
An example, where we consider an identification construction for a modified noisy Lorenz system,   can be found in section \ref{app}.

 \subsection{Proof of Theorem \ref{TNeq}}

Let us refer to the distar centers as hubs and to periphery nodes as satellites.
We suppose that satellites
do not interact each with others and a satellite interacts only with the corresponding hub.
Therefore the interaction graph resulting from the "hub disconnecting" construction
consists of $n$ disconnected distar motifs.

{\em Step  1.} Let $n=1$.  We apply  lemma \ref{Flower} to the distar graphs, see the proof of the previous theorem.
Then we have $m_1$ stable  equilibria, where $m_1$ is the number of satellites in the distar motif.

{\em Step  2}.  In the case $n >1$ we consider the disconnected interaction graph consisting of $n$
distar motifs,  where the $j$-th  distar motif
contains $m_j$ nodes. One has $m_1 + m_2 + ... + m_n=N-n$ and totally the graph consists of $N$ nodes.  For each distar we adjust the parameters
as above (see step 1).  We obtain thus $m_1 m_2 ...m_n$ of equilibria and the theorem is proven.

\section{Algorithm of construction of switchable network with prescribed dynamics} \label{app}

The proof of Theorem \ref{T3} can be used to construct practically feasible algorithms, which solve the  problem of construction of a switchable network with prescribed dynamical properties.
{As a matter of fact, we can address two different, but related problems. The first problem is the {\em synthesis} of a neural network with prescribed attractors and switchability properties. The second problem is the {\em identification} of a neural network from time series. First we state the solution of the first problem and after we describe how to resolve the second one by analogous methods.

The prescribed network properties for the {\em synthesis problem} are stated in Theorem \ref{T3}. We describe here a step by step algorithm, allowing to construct a network with these properties.}

Consider  structurally stable dynamical systems  defined by {the} equations
\begin{equation} \label{sem}
dv/dt=Q^{(l)}(v) \quad v=(v_1,...,v_n) \in \Pi(a, b) \subset {{\mathbb R}}^n,
\end{equation}
where $l=1, ...,m$ and $\Pi(a, b)$ is a  defined by \eqref{piab}. We suppose that the fields $Q^{(l)}(v)$ are sufficiently smooth, for example,
$Q^{(l)} \in C^{\infty}(\Pi(a, b))$. Without any loss of generality we can assume that
\begin{equation} \label{shift}
1 < a_i < b_i,
\end{equation}
(otherwise we can shift variables $v_i$ setting $v_i=\tilde v_i - c_i$).


{\em Step 1}. Find a sufficiently small $\epsilon$ such that perturbations of vector fields
$Q(v)^{(l)}$, which are $\epsilon$ small in $C^1$ norm, do not change topologies  of semiflows
defined by \ref{sem}.  Actually, it is hard to
compute such a value of  $\epsilon$,  so, in practice  we simply  {choose
  a small $\epsilon$ by the trial and error method.

{\em Step 2.} We find a vector field $Q(v,z)$ with $n+1$ components, where $z=v_{n+1}  \in [a_{n+1},b_{n+1}]\subset {{\mathbb R}}$ such that  the first $n$ components of $Q(v,z)$ are
defined by relations \eqref{qvz} and the $n+1$ component is defined by \eqref{qvz1}.
Let  $D= \Pi(a, b) \times [a_{n+1}, b_{n+1}]$.

To describe the next steps, first
let us introduce the functions
\begin{equation}  \label{Fjg}
G_j( \bar v, {\bf P}) =\sum_{i=1}^{N} \bar  A_{ji}\sigma(  {\bf B}_i \bar v  -  h_i),
\end{equation}
where the parameter ${\bf P}=\{N, \bar { A}_{ji}, B_{ik}, h_j, j=1,...,n+1, i,k=1,..., N \}$ and $\bar v=( v_1 ,..., v_n, z)$.

Let us observe that dynamical systems $dq/dt=Q(q)$ and $dq/dt=\gamma Q(q)$ with $\gamma >0$ have the same trajectories, invariant  sets and attractors, therefore, instead of
$Q$ we can use $\gamma Q$.
We choose  a $\gamma>0$ and a small positive $\delta < 1$  such that
\begin{equation} \label{estu1A}
 -\delta < \gamma Q_i(\bar v) < \delta,\quad \bar v \in D, \quad i=1,..., n+1
\end{equation}
and
\begin{equation}
a_i >  \delta/\lambda_i,  \quad b_i <  (1-\delta)/\lambda_i \quad i=1, ...,n+1
\label{estu2A}
\end{equation}
for $\lambda_i >1$.

Then (\ref{estu1A}) and  (\ref{estu2A}) imply that
\begin{equation} \label{estu2}
0 < \gamma Q_j(\bar v)  + \lambda_j \bar v_j < 1, \quad \bar v \in D, \ j=1,...,n+1.
\end{equation}
  Let $\sigma^{-1}$ be the function inverse to $\sigma$.  Due to (\ref{estu2}) the functions
\begin{equation} \label{Ujv}
R_j(\bar v)=\sigma^{-1} (\gamma Q_j(\bar v)  + \lambda_j \bar v_j )
\end{equation}
are correctly defined and smooth on $D$.

Now we solve the following approximation problem.

{\em To find the number $N$, the matrices $\bar {\bf A}, {\bf B}$ and vector $h$ such that
\begin{equation} \label{Approx1S}
|R_j(\bar v) - G_j( \bar v, {\bf P})| + |D_{\bar v}(R_j(\bar v) - G_j( \bar v, {\bf P}))| \le \epsilon/2, \quad j=1,..., n+1.
\end{equation}}

This problem can be resolved by standard algorithms, which perform approximations of functions  by multilayered perceptrons \cite{Barron}.  Note
that these standard methods are based on iteration procedures, which can use a large running time.

We describe here a new  variant
of the algorithm for this approximation problem, which uses a {wavelet-like} approach. This approach
does not exploit {any} iteration procedures {or} linear system solving.  All the procedure reduces to a computation  of the Fourier  and wavelet coefficients.
However, this algorithm is numerically effective only for sufficiently smooth $R_j$ with fast decreasing Fourier coefficients and for not too large dimensions $n$.

The  solution of {the} approximation problem (\ref{Approx1S}) proceeds in the two steps.

{\em Step 3}. We reduce the $n+1$-dimensional problem (\ref{Approx1S}) to a set of one-dimensional ones as follows.
Let us  approximate
the  functions $R_j$  by the Fourier expansion:
\begin{equation} \label{Four3}
  \sup_{\bar v \in D} ( |{R_j}(\bar v) -   \hat R_j(\bar v)|  + |\nabla_{\bar v}({R_j}(\bar v) -  \hat R_j(\bar v))| ) < \epsilon/4,
\end{equation}
where
\begin{equation} \label{Four4}
\hat R_j(\bar v)=  \sum_{k \in K_D}  \hat R_j(k) \exp(i (k, \bar v)),
\end{equation}
$(k, \bar v)=k_1 \bar v_1 + k_2 v_2  + ... + k_{n+1} \bar v_{n+1}$ and
the set $K_D$ of vectors $k$ is a finite subset of the $(n+1)$- dimensional lattice $L_D$
\begin{equation} \label{KD}
K_D \subset L_D=\{ k=(k_1,..., k_{n+1}):  k_i=(a_i - b_i)^{-1} \pi m_i \  for  \ some \ m_i \in {\mathbb Z} \}.
\end{equation}
The Fourier coefficients $ \hat R_j(k) $ can be computed by
$$
\hat R_j(k) =(volume(D))^{-1}  \int_{D} R_j(\bar v) \exp(-i (k, \bar v)) d\bar v.
$$
    In order to satisfy (\ref{Four3}), we take a sequence of extending
sets $K_D$.  For some $K_D$ relation (\ref{Four3}) will be satisfied  because the  Fourier coefficients $\hat R_j(k)$  fastly decrease  in $|k|$.

{\em Step 4}. We exploit the fact that the
 problem (\ref {Approx1S}) is linear with respect to the coefficients $\bar A_{ij}$.  For each $k \in K_D$ we resolve  the following one-dimensional problem. Let
\begin{equation}  \label{Fj1D}
g(q, M, a, \beta, \bar h) =\sum_{i=1}^{M} a_{i} \sigma( \beta_i ( q -  \bar h_i)).
\end{equation}

We are seeking for  integer $M>0$ and the vectors  $a=(a_1,...,a_M)$, $\beta=(\beta_1, ..., \beta_M)$ and $\bar h=(\bar h_1,..., \bar h_M)$ such that
\begin{equation} \label{Four31}
\sup_{q \in I_k}   |W_{j,k}(q) -  g(q, M, a, \beta, \bar h)|   < \epsilon(10|K_D|)^{-1},
\end{equation}

\begin{equation} \label{Four32}
\sup_{q \in I_k}   |dW_{j,k}(q)/dq -  g'(q, M, a, \beta, \bar h)|  < \epsilon_1 \le \epsilon(10|K_D|)^{-1},
\end{equation}
where  $|K_D|$ is the number of the elements $k$ in the set $K_D$,
$$
 W_{j,k}(q)=\hat R_j(k) \exp(i q),
$$
\begin{equation}  \label{Fj2D}
g'(q, M, a, \beta, \bar h) =\sum_{i=1}^{M} a_{i}\sigma^{'}( \beta_i ( q -  \bar h_i)),
\end{equation}
and $q=(k, \bar v) \in I_k$, where {$I_k$ is the interval $[q_{-}(k), q_{+}(k)]$ with}
$$
q_{-}(k)=\min_{\bar v \in D} (k, \bar  v), \quad q_{+}(k)=\min_{\bar v \in D} (k, \bar  v).
$$
These approximation problems are indexed by $(j, k)$, where $j=1,..., n+1$ and $k \in K_D$   (we temporarily omit dependence on $(j,k)$ in
$a, \beta, \bar h, M$ to simplify notation).

To resolve these  one-dimensional approximation problems, we apply a method based on the wavelet theory.  Notice that this method is numerically effective.
First  we observe that if (\ref{Four32}) is fulfilled with a sufficiently small $\epsilon_1$, then, to satisfy (\ref{Four31}), it is sufficient to add a constant term
of the form $a_{M+1} \sigma(b_{M+1} q)$ with $b_{M+1}=0$ to the sum in the right hand side of (\ref{Fj1D}).

Let us  define
the function $\psi$ by
\begin{equation} \label{psiq}
\psi(q)= \sigma^{'} (q)- \sigma^{'} (q-1).
 \end{equation}
We observe that
\begin{equation}
\int_{-\infty}^{\infty} \psi(q)dq=0
 \end{equation}
and $\psi(q) \to 0$ as $|q| \to \infty$, therefore, $\psi$ is a {wavelet-like} function.

Let us introduce the following family of functions indexed by the real parameters $r, h$:
 \begin{equation} \label{ahpsi}
\psi_{r, \xi}(q)=|r|^{-1/2} \psi(r^{-1}(q- \xi)).
 \end{equation}
 For any $f \in L_2({{\mathbb R}})$ we define  the wavelet coefficients  $T_f(r, \xi)$ of the function $f$ by
 \begin{equation} \label{wavc}
T_f (r, \xi)=\langle f, \psi_{r, \xi} \rangle = \int_{-\infty}^{\infty}  dq f(q) \psi_{r, \xi}(q).
 \end{equation}
For any smooth function $f$ with a finite support $I_R=(-R, R)$ one has the following fundamental relation:
\begin{equation}
\label{Awave}
f=c_{\psi} \int_{0}^{\infty} \int_{-\infty}^{\infty} r^{-2} dr d\xi T_f (r, \xi)\psi_{r, \xi}= f_{wav}.
 \end{equation}
for some constant $c_{\psi}$.
This equality holds in a weak sense: the left hand side and the right hand side define the same linear functionals
on $L_2({\mathbb R})$, i.e., for each smooth, well localized $g$ one has
$$
\langle f, g \rangle = \langle f_{wav}, g \rangle.
$$
Let $\delta(\epsilon) << \epsilon$ be a  small positive number.
According   to  (\ref{Awave}) we can find  positive integers $p_1$, $p_2$, points $r_1, ...,  r_{p_1}$, $\xi_1, ..., \xi_{p_2}$ and a constant $\bar c_{\psi}$ such that
the integral in the right hand side of  (\ref{Awave}) can be approximated by a finite sum:
\begin{equation}
\label{Awave2}
\sup |f(q) - \bar f_{wav}(q)| < \delta,
 \end{equation}
where
$$
\bar f_{wav}=\bar c_{\psi} \sum_{l_1=1}^{p_1}  \sum_{l_2=1}^{p_2} r_{l_1}^{-2}  T_f (r_{l_1}, \xi_{l_2})\psi_{r_{l_1}, \xi_{l_2}}.
$$
In our case for each $(j,k)$ we set $f=W_{j,k}(q)$ for $q \in I_k$ and $f=0$ for $q \notin I_k$.  We can take $r_{l_1}=r_{+}l_1/p_1$, where $r_+$ is large enough, and
$\xi_{l_2}=q_{\min} + (q_{\max} - q_{min}) l_2/p_2$, where $q_{\min} <  q_{-}(k)$, $q_{\max} > q_{+}(k)$ are sufficiently large and
$l_1=1,..., p_1, l_2=1,..., p_2$. We can renumerate the points
$(r_{l_1}, \xi_{l_2})$ by a single index $l=1,..., p$, where $p=p_1 p_2$, that gives us  $r_{l}$, $\xi_{l}$ and the wavelet coefficients $T_{l}=\bar c_{\psi} T_f (r_{l}, \xi_{l})$.

Having  $p$, $r_{l}$, $\xi_{l}$ and the wavelet coefficients $T_l$,    we obtain the following solution of the  approximation problem
(\ref{Fj1D}):
$$
M(j,k)=p,     \quad \bar h_{2l-1}(j,k)=r_l^{-1} \xi_l,  \quad  \bar h_{2l}(j,k)=r_l^{-1} (\xi_l+1),
$$
$$
\beta_{2l-1}(j,k)=\beta_{2l}(j,k)=r_l^{-1},   \quad a_{2l-1}(j,k)=-a_{2l}(j,k)=T_l,
$$
where we have introduced the index $(j,k)$ in notation for  the solution $(M, a, \beta, \bar h)$  to emphasize that  problem (\ref{Fj1D}) depends on
this index.

Finally, in the end of this step we obtain the coefficients
\begin{equation} \label{mjk}
M(j,k), a_1(j,k), ..., a_{M(j,k)}(j,k), \beta_1(j,k), ..., \beta_{M(j,k)}(j,k), \bar h_1(j,k), ..., \bar h_{M(j,k)}(j,k).
\end{equation}

{\em Step 5.}
We construct  a network with $n+1$ centers $\bar v_1,..., \bar  v_{n+1}$ and $N$ satellites as follows.  Let
${\bf C}=0$ and ${\bf D}=0$, i.e., we assume that the satellites don't interact among themselves and there are no direct interactions between the centers.
The number of satellites
is defined by
$$
N=\sum_{j=1}^{n+1} \sum_{k \in K_D} M(j,k).
$$
Each satellite can be equipped with a triple index
 $(i, j, k)$, where $j=1,...n+1$, $k \in K_D$ and $i \in \{1, ... M(j,k)\}$.  We set that all $h_j=0$, $\tilde \lambda_i=1$,
and $\lambda_j$ are chosen as above. The threshold $h_{i,j,k}$ for the satellite with the index $(i,j,k)$ is defined by
$$
h_{i,j,k}=\bar h_i(j,k)
$$
where $\bar h_i(j,k)$ are obtained at the Step 4 (see (\ref{mjk})).

Furthermore,  we define the matrices $\bar {\bf A}$ and ${\bf B}$ as follows.  One has
$$
B_{(i, j,k), l}= \beta_i(j,k) k_l,
$$
(this relation describes an action of the $l$-th center on the satellite with index $(i,j,k)$)
and
$$
\bar A_{l, (i, j,k)}=a_l(j,k)
$$
(this relation describes an action of the $l$-th center on the satellite with index $(i,j,k)$).
Here $i \in \{1, ... M(j,k)\}$, $j,l=1,..., n+1$ and $k \in K_D$.
\vspace{0.2cm}

{\bf Remark}. This algorithm can be simplified if instead networks  (\ref{cn1}), (\ref{cn2}) we use analogous networks where
satellites act on centers in a linear way:
\begin{equation}
\frac{dw_i}{dt} =
 \sigma\left(  {\bf B}_i v  + {\bf C}_i w - \tilde h_i\right) - \kappa^{-1} \tilde \lambda_i w_i,
\label{cn1L}
\end{equation}
\begin{equation}
\frac{dv_j}{dt} =
\left({\bf A}_j w  -  h_j\right) -  \lambda_j v_j,
\label{cn2L}
\end{equation}
where  $i=1,..., N_1, \ j=1,..., n$,
and the fields $Q^{(l)}$  are defined by polynomials (note that Jackson's theorems \cite{achieser2013theory} guarantee that any
 $Q$ can be approximated   by a polynomial field on  $\Pi(a, b)$ in $C^1$-norm).
Then we can simplify Step 3 and Step 4 of the algorithm as follows.  We observe that we can set $\gamma=1$ and
in this case  the functions $R_j$  have the form
\begin{equation} \label{Ujv2}
R_j(\bar v)=  Q_j(\bar v)  + \lambda_j \bar v_j.
\end{equation}
On Step 3  for polynomial functions $R_j(v)$ we  can also use simple algebraic transformations,  instead of the Fourier decomposition, to reduce the multidimensional approximation problem to one  dimensional ones.
On step 4 the function $\psi$ defined by \eqref{psiq} is well localized and therefore alternatively step 4 can be realized by standard programs using radial basic functions and  the method of least squares
(see an example on the Lorenz system below).
\vspace{0.2cm}


Let us turn now to the problem of {\em identification} of a neural network from time series produced by a  dynamical system
$dv/dt=Q(v, {\bf P})$, $v \in {\mathbb R}^n$ with unknown parameters {\bf P}. Assume that we observe a time series
$v(t_1), v(t_2),..., v(t_K)$ and the time interval between observations is small: $t_{i+1}- t_i=\Delta t << 1$.
 We want to construct a network with $n$ centers, which produces, in a sense,  analogous time series. According to (\ref{approxNN}),
 a suitable criterion of trajectory similarity is as follows. We can approximate the averages $S_{Q, \phi}$ from (\ref{SFi}) by the time series
\begin{equation} \label{SFia}
S_{Q, {\bf P}, \phi} \approx K^{-1}  \Delta T  \sum_{k=1}^K \phi({v}(t_k))=S_{Q, {\bf P}, \phi}^{(K)}.
\end{equation}
 Then, if the {network} identification is correct,
 the averages defined by time series and the corresponding ones generated by the approximating centralized neural network, should be close for smooth weight functions $\phi$:
\begin{equation} \label{approxNNa}
|S_{Q, {\bf P},  \phi}^{(K)} - S_{{G_{anN}}, \phi}^{(K)}|=Err_{approx} < \delta(\phi) << 1,
\end{equation}
{where $G_{anN}$ is the approximation of $Q$ by the neural network.}

As a first step,  we  can
approximate the unknown field $Q(v)$ by finite differences, for example, using the relation
\begin{equation} \label{semident}
Q(\tilde v_i, {\bf P})= (v(t_{i+1}) -  v(t_i)) \Delta t^{-1},  \quad \tilde v_i=(v(t_{i+1}) +  v(t_i))/2.
\end{equation}
For other values $v$ the field $Q$ can be reconstructed,  for example, by a linear interpolation.
The neural network approximation of $Q$ can be obtained by applying
the steps 2-5 of the synthesis algorithm described above.

We end this section with an illustration of the simplified variant of the identification and synthesis algorithm, see the preceding Remark.

As an example, we describe a solution of the following identification problem.  Consider time series generated by the Lorenz system perturbed by noise.  The Lorenz
system involves a {\em controller} parameter.  Adjusting the values of this parameter, we can obtain chaotic dynamics, time periodic one or dynamics with convergent trajectories.
   We are going to find a centralized network, which also has a controller parameter and can generate all this rich variety of  {trajectories}.  For chaotic and periodic trajectories
this neural approximation should exhibit dynamics
with analogous ergodic properties (in the sense of (\ref{approxNNa}).

Recall  that the Lorenz system has the form
\begin{equation}
\label{A.12}   dx/dt=\alpha(y-x), \quad dy/dt=x(\rho -z) -y,
\quad dz/dt=xy - \beta z.
 \end{equation}
This system shows a chaotic behaviour for $\alpha=10, \beta=8/3$ and $\rho=28$.
For $\alpha=10, \beta=8/3$ and $\rho \in (0,1)$  this system has a globally attracting rest point.

 We introduce  new variables $v_1=x, v_2=y, v_3=z$ and $v_4=\rho$ and  consider a more complicated modified Lorenz system with a controller parameter:
 (compare with the proof of Theorem \ref{T3}):
\begin{equation}
\label{A.13}   dv_1/dt=\alpha(v_2-v_1)=f_1, \quad dv_2/dt=r_1 v_1(v_4 -v_3) - r_2 v_2=f_2,
\end{equation}
\begin{equation} \label{A.13b}
dv_3/dt=r_3 v_1 v_2 - \beta z=f_3,  \quad dv_4/dt=\sigma_H(v_4, b_0,  h_0) - \xi v_4=f_4,
 \end{equation}
where $\sigma_H$ is  a regularized step function defined by
$H_1(w)=(1 + \exp(-b_0(w-h_0))^{-1}$
 with $b_0 >> 1$ and $ h_0=1$. We set $\xi=0.5$, $r_1=14, r_2=1, r_3=1$.  The initial data for the fourth component $v_0=v_4(0)$  is a controller parameter.
 For large $b_0$ the differential equation for $v_4$ has two stable equilibria:  $v_4^{-} \approx 0$ and $v_4^{+} \approx 2$.
Therefore,  for $v_0 \in (0, 1) $ system (\ref{A.13}),  (\ref{A.13b}) has a globally attracting rest point and for $v_0 > 1$ the  attractor of  this system  is chaotic Lorenz
one.  The parameters of this system are ${\bf P}=(\alpha, \beta, r_1, r_2, r_3)$.

Suppose we observe trajectories $v(t)$, $t \in [0, T]$ of  system (\ref{A.13})
 at some time moments $t_0=0,t_1=dt, ...,t_p=p \Delta t$.
In order to simulate experimental errors  we have perturbed the system with additive noise.
We are going to find a centralized network, which has an attractor with, in a sense, similar statistical characteristics.   More precisely, we aim to minimize $Err_{approx}$
from  relation   (\ref{approxNNa}).  For identification procedure we use a centralized network with $4$ centers $v_1, v_2, v_3$ and $v_4$. In this case steps 3, 4 can be {simplified} if we use this specific form of the modified Lorenz system.
The last center $v_4$ serves as a controller.

\vspace{0.2cm}

We state the  algorithm for the modified Lorenz system, however, the method is general and feasible  for identification by trajectories generated by all
low-dimensional dynamical systems defined by polynomial vector fields.

First we  set
\begin{equation} \label{Mat}
{\bf C}={\bf D}=0.
 \end{equation}
This means that only satellites act on centers and vice versa.  To find the matrices ${\bf A}$, ${\bf B}$ and the thresholds $h_i$, we solve the following approximation problems:
\begin{equation} \label{Approx1}
R({\bf A}, {\bf B}, h) \to min,  \quad  R=\sum_{i=1}^4 \sum_{j=1}^{p} (Q_i(t_j)- S_i({v}(t_j, {\bf A}, {\bf B}, h))^2
\end{equation}
where
\begin{equation} \label{Approx2}
Q_i(t_j)=(v_i(t_j+ \Delta t) - v_i(t_j))/\Delta t, \quad S_i({v}, {\bf A}, {\bf B}, h)=\sum_{k=1}^{N_i} A_{ik} \sigma(\sum_{j=1} B_{kj} v_j - h_{ik}).
\end{equation}

This approximation problem is nonlinear with respect to $B$ and $h$.  We can simplify this problem by the following heuristic  method.
Each function $f_i({\bf v})$ defined on a open bounded domain can be represented as a linear combination of functions $g_{l}(\vect{v} \cdot \vect{k}_{li})$,
where vectors ${\bf k}_{li}$  belong to a finite set of vectors  $K_i$.   For example, for system (\ref{A.13}),  (\ref{A.13b})
 the components $f_j$ for  $j=1,2,3$ can be represented as linear combinations of monomials:
\begin{equation}
\label{monom}
f_j(v) = g_j(v) - \lambda_j v_j, \quad  g_j(v)=\sum_{l=1}^{11} C(j,l) T_l(v)
 \end{equation}
where
$$
T_l=v_l,  \quad  l=1,2,3,4
$$
$$
T_{2l+1}=(v_1 + v_l)^2,  \quad T_{2l+2}=(v_1 - v_l)^2,   \quad l=2,3,4, \quad   T_{11}=1.
$$
and
$
\lambda_1=\alpha, \quad \lambda_2=1,   \quad \lambda_3=\beta.
$
Therefore,  $K_1=\{{\bf k}_{11}=(1, 0, 0, 0)\}$,  $K_2=\{{\bf k}_{12}=(1, 0, 1, 0),  {\bf k}_{22}=(1, 0, -1, 0), {\bf k}_{32}=(1, 0, 0, 1), {\bf k}_{42}=(1, 0, 0, -1)\}$,
  $K_3=\{ {\bf k}_{13}=(1, 1, 0, 0),  {\bf k}_{23}=(1, -1, 0, 0) \}$,
$K_4=\{ {\bf k}_{14}=(1, 0, 0, 0)$.   Let $n_i$ be the number of the vectors contained in the set $K_i$,  $n_1=1, n_2=4$, $n_3=2$ and $n_4=1$.
In this case of the modified Lorenz system, the set $K_D$  from (\ref{KD}) is the union of sets $K_i$, $i=1,...,4$.

We take a sufficiently large $N_L$, a large $b_0$ and define  the auxiliary thresholds $\bar h_{{\bf k}_{li}, j}$, where $j=1, ..., N_L$,  by
$$
 \bar h_{{\bf k}_{li}, j}=\min_{s=1,..., p , l \in K_i} {v}(t_s) \cdot \vect{k}_{li}  + j
(\max_{s=1,..., p , l \in K_i} {v}(t_s) \cdot \vect{k}_{li}-\min_{s=1,..., p , l \in K_i} {v}(t_s) \cdot \vect{k}_{li})/NL.
$$
We  seek  coefficients $\bar A_{il, {\bf k}_{li}}$ and $C_i$, which minimize  $R_i(\bar {\bf A}, C_i)$ for $i=1,2,3, 4$:

\begin{equation} \label{Approx1SS}
R_i(\bar {\bf A}, C_i) \to min,  \quad  R_i= \sum_{j=1}^{p} (Q_i(t_J)- \tilde S_i({v}(t_j), \bar{\bf A}, C_i ))^2
\end{equation}

where
\begin{equation} \label{Approx2SS}
\tilde S_i({v}, \bar {\bf A}, C)=C_i+\sum_{l=1}^{n_i} \sum_{j=1}^{N_L}\bar A_{ij, {\bf k}_{li}} \sigma( b_0({\bf k}_{li} \cdot {v} - \bar h_{{\bf k}_{li},j})).
\end{equation}
Note that since $\tilde S_i$ are linear functions of $\bar A_{il, {\bf k}_{li}}$ and $C_i$,  problems (\ref{Approx1SS}) can be solved by the least square method.
The important advantage of this approach is that approximations can be done independently for different components $i$.

This approximation produces a centralized network involving $4$ centers and $N=8 N_L  + 8$ satellites.
Indeed,  each vector ${\bf k}_{li}$ associated with a quadratic term $T_l$,  gives us $N_L$ sattellites to approximate this term.  Moreover,  we use  $4$ satellites for approximations
of the linear terms and
4 satellites are necessary for  constants $C_i$ in the right hand sides of (\ref{Approx2SS}).
\vspace{0.2cm}

\begin{figure}[h!]
\centerline{
\includegraphics[width=80mm]{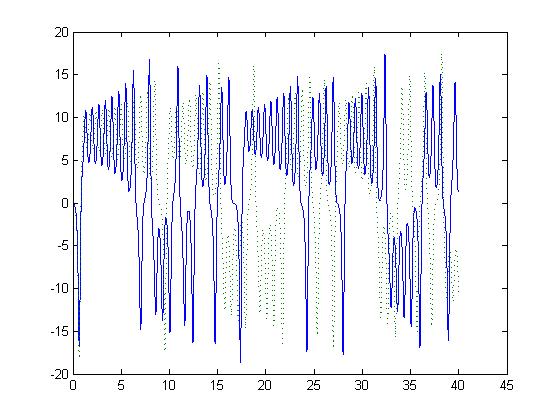}}
\caption{\small
This plot shows trajectories of $v_1$-component of
the Lorenz system perturbed by noise (the solid curve) and its neural approximation with $N=20$ satellites (the dotted curve). The curves are not close but they
exhibit almost identical statistical properties  ($Err_{approx}=0.008$  (the white noise level is $0.05$, solutions have been obtained by the Euler method with the time step $0.001$ on the interval
$[0,40])$.
 }
\label{fig:4}
\end{figure}

The numerical simulations give the following results. The trajectories to identify are produced by the Euler method applied to the system (\ref{A.13}),  (\ref{A.13b}) perturbed by  noise, where
the time step $0.005$ on the interval $[0, 50]$, the noise is simulated by $\epsilon_N \omega(t_i)$, where  $\omega(t)$ is the standard white noise and
$\epsilon_N=0.05$.      As a result of minimization procedure, we have obtained the errors $R_i$ of the order $0.01- 0.1$.  The trajectories of the system (\ref{A.13}),  (\ref{A.13b}) perturbed by noise and the corresponding neural networks are not close but they have a similar form and statistical characteristics
that is confirmed by the value $Err_{approx} $ (defined by  (\ref{approxNNa})), which is $0.008$,
where the test function $\phi$ is $\phi({v})=v_1^2 + v_2^2/2 - 2v_3$.
These results are illustrated by Fig. 4.


\section{Conclusion and discussion}

In this paper, we have proposed a complete analytic theory of maximally flexible and switchable Hopfield networks.
 We shown that dynamics of a network with $n$ slow components $v_1, ..., v_n$ can be reduced to a system of $n$ differential equations defined by
a smooth $n$ dimensional vector field $F(v)$.
 If these slow components are hubs, i.e.,  they are  connected with a number of other weakly connected nodes (satellites) and center-satellite interactions dominate
inter-satellite  forces, then  the network becomes maximally flexible. Namely,  by adjusting only center-satellite interactions we can
obtain smooth $F$ of arbitrary forms.

These networks are also maximally switchable.  We describe networks of a special architecture, which
contains a controller hub. By changing the state of this hub and
the hub response time parameter $\xi$ one can completely change the network dynamics
from an unique global attractive steady state to any combination of periodic or chaotic
attractors.

Our results provide a rigorous framework for the idea that centralized networks are flexible. We also
propose  mechanisms for switching between attractors of these networks with controller
hubs. In functional genomics there are numerous examples when transitions between attractors of gene regulatory networks
can be triggered by controller proteins having
multiple states sometimes resulting from interactions
with micro-RNA satellites \cite{carthew2006gene}. Similarly, neurons having
multiple internal states can trigger
phase transitions of brain networks suggesting that single neuron activation could be used
for neural network control \cite{fujisawa2006single}.

The proofs of our results are constructive and are based on an algorithm allowing the network
reconstruction. This algorithm has several potential applications in biology.
Identified networks can be used to study emergent network properties such as robustness,
controllability and switchability. Gene networks
with the desired switchability properties could be build by synthetic biology tools for various
applications in biotechnology. Furthermore, maximal switchable network models
can be  used in neuroscience to relate structure and function in the brain activity,
or in genetics to explain how a minimal number of mutations
can induce large phenotypic changes from one type of adaptive behavior to another one.


{\bf Acknowledgements}

S.V. was financially supported by Government of Russian Federation, Grant 074-U01, also supported in part by grant
RO1 OD010936 (formerly RR07801) from the US NIH and by grant -а of Russian Fund of Basic Research.
O.R. was supported by the  Labex EPIGENMED (ANR-10-LABX-12-01).
{The authors are grateful to the anonymous referees for their useful remarks, that helped improve the text}.


\begin{thebibliography}{10}

\bibitem{achieser2013theory}
Naum~I Achieser.
\newblock {\em {Theory of approximation}}.
\newblock Courier Corporation, 2013.

\bibitem{AB}
R{\'e}ka Albert and Albert-L{\'a}szl{\'o} Barab{\'a}si.
\newblock {Statistical mechanics of complex networks}.
\newblock {\em Reviews of modern physics}, 74(1):47, 2002.

\bibitem{albert2000error}
R{\'e}ka Albert, Hawoong Jeong, and Albert-L{\'a}szl{\'o} Barab{\'a}si.
\newblock {Error and attack tolerance of complex networks}.
\newblock {\em Nature}, 406(6794):378--382, 2000.

\bibitem{bar2004response}
Yaneer Bar-Yam and Irving~R Epstein.
\newblock {Response of complex networks to stimuli}.
\newblock {\em Proceedings of the National Academy of Sciences of the United
  States of America}, 101(13):4341--4345, 2004.

\bibitem{Barron}
Andrew~R Barron.
\newblock {Universal approximation bounds for superpositions of a sigmoidal
  function}.
\newblock {\em Information Theory, IEEE Transactions on}, 39(3):930--945, 1993.

\bibitem{Bas1}
Jordi Bascompte.
\newblock {Networks in ecology}.
\newblock {\em Basic and Applied Ecology}, 8(6):485--490, 2007.

\bibitem{carlson2002complexity}
Jean~M Carlson and John Doyle.
\newblock {Complexity and robustness}.
\newblock {\em Proceedings of the National Academy of Sciences}, 99(suppl
  1):2538--2545, 2002.

\bibitem{carthew2006gene}
Richard~W Carthew.
\newblock {Gene regulation by microRNAs}.
\newblock {\em Current opinion in genetics \& development}, 16(2):203--208,
  2006.

\bibitem{chialvo}
Dante~R Chialvo.
\newblock {Emergent complex neural dynamics}.
\newblock {\em Nature physics}, 6(10):744--750, 2010.

\bibitem{cohen2001breakdown}
Reuven Cohen, Keren Erez, Daniel Ben-Avraham, and Shlomo Havlin.
\newblock {Breakdown of the Internet under intentional attack}.
\newblock {\em Physical review letters}, 86(16):3682, 2001.

\bibitem{cornelius2013realistic}
Sean~P Cornelius, William~L Kath, and Adilson~E Motter.
\newblock {Realistic control of network dynamics}.
\newblock {\em Nature communications}, 4, 2013.

\bibitem{cowan2012nodal}
Noah~J Cowan, Erick~J Chastain, Daril~A Vilhena, James~S Freudenberg, and
  Carl~T Bergstrom.
\newblock {Nodal dynamics, not degree distributions, determine the structural
  controllability of complex networks}.
\newblock {\em PloS one}, 7(6):e38398, 2012.

\bibitem{Deco}
Gustavo Deco and Viktor~K Jirsa.
\newblock {Ongoing cortical activity at rest: criticality, multistability, and
  ghost attractors}.
\newblock {\em The Journal of Neuroscience}, 32(10):3366--3375, 2012.

\bibitem{edwards1999parkinsonian}
Roderick Edwards, Anne Beuter, and Leon Glass.
\newblock {Parkinsonian tremor and simplification in network dynamics}.
\newblock {\em Bulletin of mathematical biology}, 61(1):157--177, 1999.

\bibitem{fujisawa2006single}
Shigeyoshi Fujisawa, Norio Matsuki, and Yuji Ikegaya.
\newblock {Single neurons can induce phase transitions of cortical recurrent
  networks with multiple internal states}.
\newblock {\em Cerebral Cortex}, 16(5):639--654, 2006.

\bibitem{gao2014target}
Jianxi Gao, Yang-Yu Liu, Raissa~M D'Souza, and Albert-L{\'a}szl{\'o}
  Barab{\'a}si.
\newblock {Target control of complex networks}.
\newblock {\em Nature communications}, 5, 2014.

\bibitem{Hale}
Jack~K Hale.
\newblock {\em {Asymptotic behavior of dissipative systems}}, volume~25.
\newblock American Mathematical Soc., 2010.

\bibitem{He}
Dan Henry.
\newblock {\em {Geometric theory of semilinear parabolic equations}}, volume
  840.
\newblock Springer-Verlag, Berlin, 1981.

\bibitem{Hopfield}
John~J Hopfield.
\newblock {Neural networks and physical systems with emergent collective
  computational abilities}.
\newblock {\em Proceedings of the National Academy of Sciences},
  79(8):2554--2558, 1982.

\bibitem{hopfield1986computing}
John~J Hopfield, David~W Tank, et~al.
\newblock Computing with neural circuits- a model.
\newblock {\em Science}, 233(4764):625--633, 1986.

\bibitem{Huang2009869}
Sui Huang, Ingemar Ernberg, and Stuart Kauffman.
\newblock {Cancer attractors: A systems view of tumors from a gene network
  dynamics and developmental perspective}.
\newblock {\em {Seminars in Cell $\&$ Developmental Biology }}, 20(7):869 --
  876, 2009.

\bibitem{Jeong2}
Hawoong Jeong, Sean~P Mason, A-L Barab{\'a}si, and Zoltan~N Oltvai.
\newblock {Lethality and centrality in protein networks}.
\newblock {\em Nature}, 411(6833):41--42, 2001.

\bibitem{Jeong1}
Hawoong Jeong, B{\'a}lint Tombor, R{\'e}ka Albert, Zoltan~N Oltvai, and A-L
  Barab{\'a}si.
\newblock {The large-scale organization of metabolic networks}.
\newblock {\em Nature}, 407(6804):651--654, 2000.

\bibitem{jia2013control}
Tao Jia and Albert-L{\'a}szl{\'o} Barab{\'a}si.
\newblock {Control capacity and a random sampling method in exploring
  controllability of complex networks}.
\newblock {\em Scientific reports}, 3, 2013.

\bibitem{Lip}
Angelo~Valleriani J{\"o}rg~Menche and Reinhard Lipowsky.
\newblock {Dynamical processes on dissortative scale-free networks}.
\newblock {\em EPL (Europhysics Letters)}, 89(1):18002, 2010.

\bibitem{kifer1986general}
Yuri Kifer.
\newblock {General random perturbations of hyperbolic and expanding
  transformations}.
\newblock {\em Journal d’Analyse Math{\'e}matique}, 47(1):111--150, 1986.

\bibitem{lai2014controlling}
Ying-Cheng Lai.
\newblock {Controlling complex, non-linear dynamical networks}.
\newblock {\em National Science Review}, 1(3):339--341, 2014.

\bibitem{Li}
Xin Li, Justin~J Cassidy, Catherine~A Reinke, Stephen Fischboeck, and Richard~W
  Carthew.
\newblock {A microRNA imparts robustness against environmental fluctuation
  during development}.
\newblock {\em Cell}, 137(2):273--282, 2009.

\bibitem{li1989modeling}
Zhaoping Li and JJ~Hopfield.
\newblock {Modeling the olfactory bulb and its neural oscillatory processings}.
\newblock {\em Biological cybernetics}, 61(5):379--392, 1989.

\bibitem{lin1974structural}
Ching~Tai Lin.
\newblock {Structural controllability}.
\newblock {\em Automatic Control, IEEE Transactions on}, 19(3):201--208, 1974.

\bibitem{liu2011controllability}
Yang-Yu Liu, Jean-Jacques Slotine, and Albert-L{\'a}szl{\'o} Barab{\'a}si.
\newblock {Controllability of complex networks}.
\newblock {\em Nature}, 473(7346):167--173, 2011.

\bibitem{maass1991computational}
Wolfgang Maass, Georg Schnitger, and Eduardo~D Sontag.
\newblock {On the computational power of sigmoid versus Boolean threshold
  circuits}.
\newblock In {\em Foundations of Computer Science, 1991. Proceedings., 32nd
  Annual Symposium on}, pages 767--776. IEEE, 1991.

\bibitem{Rein1}
Eric Mjolsness, David~H Sharp, and John Reinitz.
\newblock {A connectionist model of development}.
\newblock {\em Journal of theoretical Biology}, 152(4):429--453, 1991.

\bibitem{motter2015networkcontrology}
Adilson~E Motter.
\newblock {Networkcontrology}.
\newblock {\em Chaos: An Interdisciplinary Journal of Nonlinear Science},
  25(9):097621, 2015.

\bibitem{nepusz2012controlling}
Tam{\'a}s Nepusz and Tam{\'a}s Vicsek.
\newblock {Controlling edge dynamics in complex networks}.
\newblock {\em Nature Physics}, 8(7):568--573, 2012.

\bibitem{newman2003structure}
Mark~EJ Newman.
\newblock {The structure and function of complex networks}.
\newblock {\em SIAM review}, 45(2):167--256, 2003.

\bibitem{orr2005genetic}
H~Allen Orr.
\newblock {The genetic theory of adaptation: a brief history}.
\newblock {\em Nature Reviews Genetics}, 6(2):119--127, 2005.

\bibitem{Temam}
B.~Nicolaenko P.~Constantin, C.~Foias and R.~Temam.
\newblock {Integral Manifolds and Inertial Manifolds for Dissipative Partial
  Differential Equations}.
\newblock {\em Springer-Verlag, Applies Mathematical Sciences Series,}, 70.

\bibitem{pan2014structural}
Yujian Pan and Xiang Li.
\newblock {Structural controllability and controlling centrality of temporal
  networks}.
\newblock {\em PloS one}, 9(4):e94998, 2014.

\bibitem{pasqualetti2014controllability}
Fabio Pasqualetti, Sandro Zampieri, and Francesco Bullo.
\newblock {Controllability metrics, limitations and algorithms for complex
  networks}.
\newblock {\em Control of Network Systems, IEEE Transactions on}, 1(1):40--52,
  2014.

\bibitem{pereira2010hub}
Tiago Pereira.
\newblock {Hub synchronization in scale-free networks}.
\newblock {\em Physical Review E}, 82(3):036201, 2010.

\bibitem{pereira2013connectivity}
Tiago Pereira, Deniz Eroglu, G~Baris Bagci, Ugur Tirnakli, and Henrik~Jeldtoft
  Jensen.
\newblock {Connectivity-driven coherence in complex networks}.
\newblock {\em Physical review letters}, 110(23):234103, 2013.

\bibitem{SyS}
Subramoniam Perumal and Ali~A Minai.
\newblock {Stable-yet-switchable (sys) attractor networks}.
\newblock In {\em Neural Networks}, pages 2509--2516, 2009.

\bibitem{P1}
Peter {Pol{\'a}{\v{c}}ik}.
\newblock {Complicated dynamics in scalar semilinear parabolic equations in
  higher space dimension}.
\newblock {\em Journal of differential equations}, 89(2):244--271, 1991.

\bibitem{Ruelle}
David Ruelle.
\newblock {\em {Elements of differentiable dynamics and bifurcation theory}}.
\newblock Elsevier, 2014.

\bibitem{ruths2014control}
Justin Ruths and Derek Ruths.
\newblock {Control profiles of complex networks}.
\newblock {\em Science}, 343(6177):1373--1376, 2014.

\bibitem{stam1995investigation}
CJ~Stam, B~Jelles, HAM Achtereekte, SARB Rombouts, JPJ Slaets, and RWM Keunen.
\newblock {Investigation of EEG non-linearity in dementia and Parkinson's
  disease}.
\newblock {\em Electroencephalography and clinical neurophysiology},
  95(5):309--317, 1995.

\bibitem{stroud2015dynamics}
Jake Stroud, Mauricio Barahona, and Tiago Pereira.
\newblock {Dynamics of Cluster Synchronisation in Modular Networks:
  Implications for Structural and Functional Networks}.
\newblock In {\em Applications of Chaos and Nonlinear Dynamics in Science and
  Engineering-Vol. 4}, pages 107--130. Springer, 2015.

\bibitem{sun2013controllability}
Jie Sun and Adilson~E Motter.
\newblock {Controllability transition and nonlocality in network control}.
\newblock {\em Physical Review Letters}, 110(20):208701, 2013.

\bibitem{talagrand1998rigorous}
Michel Talagrand.
\newblock Rigorous results for the hopfield model with many patterns.
\newblock {\em Probability theory and related fields}, 110(2):177--275, 1998.

\bibitem{tanizawa2005optimization}
Toshi Tanizawa, Gerald Paul, Reuven Cohen, Shlomo Havlin, and H~Eugene Stanley.
\newblock {Optimization of network robustness to waves of targeted and random
  attacks}.
\newblock {\em Physical review E}, 71(4):047101, 2005.

\bibitem{Vak1}
SA~Vakulenko.
\newblock {A system of coupled oscillators can have arbitrary prescribed
  attractors}.
\newblock {\em Journal of Physics A: Mathematical and General}, 27(7):2335,
  1994.

\bibitem{Vak2}
SA~Vakulenko.
\newblock {Dissipative systems generating any structurally stable chaos}.
\newblock {\em Advances in Differential Equations}, 5(7-9):1139--1178, 2000.

\bibitem{vakulenko2011flexible}
Sergei Vakulenko and Ovidiu Radulescu.
\newblock {Flexible and robust patterning by centralized gene networks}.
\newblock {\em Fundamenta Informaticae}, 118(4):345--369, 2012.

\bibitem{vakulenko2012flexible}
Sergey~A Vakulenko and Ovidiu Radulescu.
\newblock {Flexible and robust networks}.
\newblock {\em Journal of bioinformatics and computational biology},
  10(02):1241011, 2012.

\bibitem{Viana}
M~Viana.
\newblock {Dynamics : A Probabilistic and Geometric Perspective}.
\newblock {\em Documenta Mathematica}, Extra Volume ICM:557--578, 1998.

\bibitem{vohradsky2001neural}
JI{\v{R}}{\'I} Vohradsk{\'y}.
\newblock {Neural network model of gene expression}.
\newblock {\em The FASEB Journal}, 15(3):846--854, 2001.

\bibitem{wu2014transittability}
Fang-Xiang Wu, Lin Wu, Jianxin Wang, Juan Liu, and Luonan Chen.
\newblock {Transittability of complex networks and its applications to
  regulatory biomolecular networks}.
\newblock {\em Scientific reports}, 4, 2014.

\bibitem{yan2012controlling}
Gang Yan, Jie Ren, Ying-Cheng Lai, Choy-Heng Lai, and Baowen Li.
\newblock {Controlling complex networks: How much energy is needed?}
\newblock {\em Physical review letters}, 108(21):218703, 2012.

\bibitem{young1986stochastic}
Lai-Sang Young.
\newblock {Stochastic stability of hyperbolic attractors}.
\newblock {\em Ergodic Theory and Dynamical Systems}, 6(02):311--319, 1986.

\bibitem{yuan2013exact}
Zhengzhong Yuan, Chen Zhao, Zengru Di, Wen-Xu Wang, and Ying-Cheng Lai.
\newblock {Exact controllability of complex networks}.
\newblock {\em Nature communications}, 4, 2013.

\end{thebibliography}

\end{document}